\documentclass[%
 aip,
 amsmath,amssymb,
 reprint,
]{article}

\usepackage{amsthm,amsmath,stmaryrd,bbm,hyperref,geometry,color}
\usepackage[utf8]{inputenc}
\usepackage[english]{babel}
\usepackage{graphicx}
\usepackage{amsfonts,amssymb}
\usepackage{verbatim}
\usepackage{enumitem}
\usepackage{algorithm}
\usepackage{algpseudocode}
\usepackage{multirow}
\usepackage{caption}
\usepackage{subcaption}
\usepackage{tabularx}
\usepackage[export]{adjustbox}

\usepackage{authblk}
\usepackage[table,xcdraw]{xcolor}

\newcommand{\po}{\left(}
\newcommand{\pf}{\right)}
\newcommand{\co}{\left[}
\newcommand{\cf}{\right]}
\newcommand{\cco}{\llbracket}
\newcommand{\ccf}{\rrbracket}
\newcommand{\R}{\mathbb R} 
\newcommand{\T}{\mathbb T} 
\newcommand{\Z}{\mathbb Z}
\newcommand{\PP}{\mathbb P}
\newcommand{\E}{\mathbb E} 
 
\newcommand{\N}{\mathbb N} 
\newcommand{\dd}{\text{d}}

\newcommand{\veps}{\varepsilon}

\newtheorem*{remark}{Remark}

\hypersetup{
    unicode=false,          % non-Latin characters in Acrobatï¿œs bookmarks
    pdftoolbar=true,        % show Acrobatï¿œs toolbar?
    pdfmenubar=true,        % show Acrobatï¿œs menu?
    pdffitwindow=false,     % window fit to page when opened 
    pdfstartview={FitH},    % fits the width of the page to the window
    citecolor=blue,
    colorlinks=true,       % false: boxed links; true: colored links
    linkcolor=blue,          % color of internal links (change box color with linkbordercolor)
    filecolor=blue,      % color of file links
    urlcolor=blue           % color of external links
}

\title{\begin{center}
\large
{\bf  Velocity Jumps for Molecular Dynamics}
\end{center}}
\date{}

\author[1,2,3]{Nicolaï Gouraud}
\author[1,3]{Louis Lagardère}
\author[1]{Olivier Adjoua}
\author[1]{Thomas Pl{\'e}}
\author[1,2,4,*]{Pierre Monmarch{\'e}}
\author[1,3,*]{Jean-Philip Piquemal}
\affil[1]{Sorbonne Université, CNRS, LCT UMR 7616, 75005 Paris, France}
\affil[2]{Sorbonne Université, Université Paris Cité, CNRS, INRIA, LJLL UMR 7598, 75005 Paris, France}
\affil[3]{Qubit Pharmaceuticals, Advanced Research Department, 75014 Paris, France}
\affil[4]{Institut Universitaire de France, 75005 Paris, France}
\affil[*]{pierre.monmarche@sorbonne-universite.fr,jean-philip.piquemal@sorbonne-universite.fr}

\begin{document}

\maketitle

\begin{abstract}
    We introduce the Velocity Jumps approach, denoted as JUMP, a new class of Molecular dynamics integrators, replacing the Langevin dynamics by a hybrid model combining a classical Langevin diffusion and a piecewise deterministic Markov process, where the expensive computation of long-range pairwise interactions is replaced by a resampling of the velocities at random times. This framework allows for an acceleration in the simulation speed while preserving sampling and dynamical properties such as the diffusion constant. It can also be integrated in classical multi-timestep methods, pushing further the computational speedup, while avoiding some of the resonance issues of the latter thanks to the random nature of jumps. The JUMP, JUMP-RESPA and JUMP-RESPA1 integrators have been implemented in the GPU-accelerated version of the Tinker-HP package and are shown to provide significantly enhanced performances compared to their BAOAB, BAOAB-RESPA and BAOAB-RESPA1 counterparts respectively.
\end{abstract}

\section{Introduction}

Molecular dynamics (MD)~\cite{liquids,md} is a popular numerical tool to infer macroscopic properties of matter from simulations at a microscopic level, with applications ranging from material sciences to biology~\cite{appli1,appli2,appli3}. Because of the discrepancy between the microscopic and macroscopic time and space scales, in order to obtain relevant results, one needs to simulate complex molecular systems with a high precision and for a very long time~\cite{spike,longsimu,bigsimu2,ShawmilliANTON,bigsimu3,bigsimu1}. In recent years, this has been allowed by the development of empirical force fields (such as CHARMM~\cite{CharmmFF}, AMBER~\cite{AmberFF}, OPLS~\cite{opls}, GROMOS~\cite{gromos}, AMOEBA~\cite{amoeba}, SIBFA ~\cite{SIBFA} and others~\cite{otherff,article71,article72}), as well as the rise of high-performance computing with GPUs~\cite{tinkerhpgpu,gpu2,gpu1,gpu3}, allowing to run massively parallel codes. Nowadays, many molecular simulation softwares are available (LAMMPS~\cite{lammps}, NAMD~\cite{NAMD}, GROMACS~\cite{GROMACS}, AMBER~\cite{AMBERsoftware}, Tinker-HP~\cite{tinkerhp} and others~\cite{ShawmilliANTON,DOMDEC,GENESIS,PLIMPTON19951}) and a lot of attention is drawn to improving the efficiency of the simulation algorithms.

In the framework of Langevin dynamics, multi-timestep methods~\cite{GibsonCarter,pushinglimits,TuckermanRossiBerne,multitimestep} are now a standard way to make faster simulations than classical integrators such as BAOAB~\cite{BouRabee,baoab2,stoltzsplitting,baoab1}, but they are limited by resonance effects that bound the maximum usable time step at a given accuracy~\cite{Skeelresonance1,Skeelresonance2}.
Various alternative strategies have been proposed such as the Generalized Langevin Equation (GLE) \cite{GLE} or the stochastic isokinetic extended phase-space algorithm \cite{isokin3,isokin,isokinpol}.
They yield significant acceleration but rely either on some empirical fitting \cite{GLE} or have an important negative impact on the dynamical properties such as the diffusion coefficient \cite{isokin3}, which is an indication of a limitation in the effective sampling rate. In the last case, the computational gain is reduced by the fact that a longer trajectory is necessary to keep the same amount of sampling. To speed up molecular dynamics without resorting to some fitting and while not affecting too much the dynamics~\cite{GLE, BerendsenHMR}, one possibility is to combine well-chosen integrators within a multi-split approach like for BAOAB-RESPA1 to push forward the stability limit~\cite{pushinglimits}.

Here we will focus on an alternative approach that involves replacing the Langevin dynamics by a hybrid model combining a classical Langevin diffusion and a piecewise deterministic Markov process, in which some long-range, computationally expensive force are treated in an adaptive way at random times. The initial idea of this strategy was first introduced in~\cite{velocity} in a particular case, and further studied theoretically in~\cite{splittinglangevinjumps}. This hybrid model still samples exactly from $\mu$ but can be simulated using a numerical splitting scheme that requires fewer gradient computations per time step, with a precision of the same order (in the time step) as the classical splitting schemes of the Langevin diffusion such as BAOAB. Moreover, it can be tuned to be arbitrarily close to the Langevin dynamics in terms of stochastic trajectories (see~\cite[Theorem 3.6]{MRZ}), which makes it suitable to estimate the dynamical properties of the process (with, of course, a trade-off between the accuracy of these dynamical properties and the numerical cost of the simulation, as would be with any numerical approximation of the Langevin equation). Finally, this versatile framework is parallelizable (allowing GPU implementations) and can be combined with the multi-timestep methods, pushing further the computational speedup while avoiding some of the resonance issues of the latest~\cite{Skeelresonance1,Skeelresonance2}. This new integrator has been implemented in the Tinker-HP software~\cite{tinkerhp}, both on CPU and GPU versions~\cite{tinkerhpgpu}, the latter allowing simulations on much larger systems.

The paper is organized as follows. Section~\ref{sec:nummethod} is devoted to the description of the numerical method: we define in Sections~\ref{sec:proc-continu} and~\ref{sec:Jump_mechanism} the continuous-time hybrid process, Sections~\ref{sec:splitting} and~\ref{sec:thinning} describe the simulation techniques and how they yield a computational advantage. Then, in Section~\ref{sec:application}, we see how our approach can be applied to Coulomb and van der Waals interactions in a classical force field. Finally, Section~\ref{sec:simus} is devoted to numerical experiments, both on CPU and GPU versions of the Tinker-HP software~\cite{tinkerhp}. The supporting information gives the proof of some of the results and other technical details.

\subsection{Notations}
In all the following, $|\cdot|$ denotes the Euclidean norm in $\R^{d}$ (or the Frobenius norm when it is applied to matrices), and $\cdot$ the standard dot product. For all $x\in\R$, $(x)_+ = \max(0,x)$ denotes the positive part of $x$.
\section{Numerical method}\label{sec:nummethod}

\subsection{The velocity jump Langevin process}\label{sec:proc-continu}

Let us consider a system of $N$ interacting atoms. We denote $(x,v)\in\R^{6N}$ their positions and velocities, $M = \text{diag}(m_1 I_3,\dots,m_N I_3)$ their mass matrix, $\beta = 1/(k_B T)$  the inverse temperature of the system (where $k_B$ is the Boltzmann constant), $U:\R^{3N}\rightarrow\R$ the potential energy function encoding the interactions between the particles and finally $H(x,v) =  U(x) + \frac{1}{2}v^T M v$ the Hamiltonian of the system, corresponding to its total energy.

In the canonical ensemble, the system is described by a probability measure that gives, for any given state $(x,v)\in\R^{6N}$, the probability that the system is in the configuration $(x,v)$. This probability measure is called the Boltzmann-Gibbs measure, defined by

\begin{equation}\label{eq:Gibbs-measure}
    \dd\mu(x,v) = \frac{1}{\mathcal{Z}_\mu}\exp(-\beta H(x,v))\dd x\dd v\,,
\end{equation}
where $\mathcal{Z}_\mu = \int_{\R^{6N}}\exp(-\beta H(x,v))\dd x\dd v$.
Macroscopic quantities are then described as expectations of an observable with respect to this Gibbs measure. In many cases, computing them analytically is impossible, since they are high-dimensional integrals involving an unknown constant $\mathcal{Z}_\mu$. However, it is possible to approximate $\E_\mu[\varphi]$ by simulating a long trajectory of a Markov process $(X_t,V_t)_{t\geqslant 0}$ that is ergodic with respect to $\mu$, which means that for any $\varphi$ (in some suitable class of functions), almost surely:
\[\frac{1}{t}\int_0^t \varphi(X_s,V_s)\dd s \underset{t\rightarrow\infty}\longrightarrow \int_{\R^{6N}}
\varphi(x,v)\dd \mu(x,v) = \E_\mu[\varphi]\,.\]
A popular process that has this property (under mild conditions on the potential $U$) is the Langevin dynamics, defined as the solution of the following SDE:

\begin{equation}\label{eq:langevin}
\left\{\begin{array}{rcl}
\dd X_t & = &  V_t \dd t \\
\dd V_t & = & -M^{-1}\nabla U (X_t) \dd t - \gamma V_t \dd t - M^{-1/2}\sqrt{2\gamma\beta^{-1}}\dd W_t\,,
\end{array}\right.
\end{equation}
where $(W_t)_{t\geqslant 0}$ is a standard Brownian motion in $\R^{3N}$ and $\gamma > 0$ is a friction parameter.

In practice, in many molecular simulations, periodic boundary conditions are enforced (typically to ensure bounded solutions in systems that do not have a confining potential), which means that the position lies in the periodic flat torus $\T^{3N} = \R^{3N}/\Z^{3N}$. The measure~\eqref{eq:Gibbs-measure} and the SDE~\eqref{eq:langevin} are then understood in $\T^{3N}\times\R^{3N}$.

The following general procedure was first described in~\cite{velocity}.
The first step is to decompose the forces as a sum of $K \geq 1$ vector fields $F_i$:
\[\nabla U(x) = \sum_{i=0}^K F_i(x)\,,\]
such that, typically, $F_0$ gathers the computationally inexpensive components (e.g. short-range interactions exhibiting fast variation), while the $F_i$ terms, for $i\geqslant 1$, represent longer-range forces, which are more numerically intensive than $F_0$ (as each atom interacts with all others through these forces, unlike the short-range ones).

We now introduce the velocity jump Langevin process $(X_t,V_t)_{t\geqslant0}$ as the following. The dynamics follows the Langevin diffusion associated with the force $F_0$:

\begin{equation}\label{eq:langevinbis}
\left\{\begin{array}{rcl}
\dd X_t & = &  V_t \dd t \\
\dd V_t & = & -M^{-1}F_0 (X_t) \dd t - \gamma V_t \dd t - M^{-1/2}\sqrt{2\gamma\beta^{-1}}\dd W_t\,,
\end{array} \right.
\end{equation}
with the velocity $V_t$ undergoing additional random jumps at rate $\lambda_i(x,v)$ following a jump kernel $q_i(x,v,\dd v')$ (that describes the probability distribution of the velocity after a jump), both defined below in Section~\ref{sec:Jump_mechanism}, as also explained in~\cite{velocity} in a particular case. The jump mechanism $(\lambda_i,q_i)$ depends on the force $F_i$ in a way that ensures that the equilibrium measure of the process in indeed the canonical measure~\eqref{eq:Gibbs-measure}. Informally, the process follows \eqref{eq:langevinbis} but, between times $t$ and $t+\delta$ for a small $\delta$, with probability  $\lambda_i(X_t,V_t)\delta + o(\delta)$,  its velocity $v$ is re-sampled according to the kernel $q_i(X_t,V_t,\dd v')$. It can be seen as a hybrid between a diffusion and a piecewise deterministic Markov process (PDMP) such as considered in \cite{MonmarcheRTP,PetersdeWith,DoucetPDMCMC}, which have recently drawn much attention in numerical probabilities and Bayesian statistics, see~\cite{PetersdeWith,bertazzi2021approximations,M45,ZigZag,corbella_automatic,cotter2020nuzz}. As we will explain, simulating this process in a certain way yields a computational advantage over the simulation of the Langevin diffusion, because the computation of the force $F_i$ will only be required when a jump of type $i$ is proposed, which will not happen at each time step for every $i$. We refer to the SI, Section 1, for a more rigorous definition.

\subsection{The jump mechanism}\label{sec:Jump_mechanism}

The velocity jump Langevin process defined in Section \ref{sec:proc-continu} was described in \cite{velocity} in the particular case where the jump mechanism corresponds to the Bouncy Particle Sampler (BPS) \cite{bps}, i.e. with
\begin{equation}\label{eq:tauxbps}
\lambda_i(x,v) = \beta (v\cdot F_i(x))_{+}\,,
\end{equation}
and deterministic jumps $q_i(x,v,\dd v')=\delta_{R_i(x,v)}(\dd v')$, where
\begin{equation}\label{eq:reflexion}
R_i(x,v) = v - 2 \frac{v\cdot F_i(x)}{F_i(x)\cdot  M^{-1}F_i(x)}M^{-1}F_i(x)\,.
\end{equation}
Although this jump mechanism gives good results in a certain framework~\cite{velocity}, it also suffers some limitations, that we will explain below, in Section \ref{sec:application}. We now define a more general jump mechanism based on~\cite{MRZ}, that can be seen as an interpolation between the BPS and Hamiltonian dynamics.
Let $\veps(x) > 0$ be a positive function of the position. For each $1 \leq i \leq K$,  We define the following jump rate:
\begin{equation}\label{eq:tauxsaut}
\lambda_i(x,v) = \frac{\sqrt{\beta}}{\veps(x)}|M^{-1/2}F_i(x)|\Theta(\eta)\,,
\end{equation}
where \[ \eta =\veps(x)\sqrt{\beta}\frac{v \cdot F_i(x)\,,}{|M^{-1/2}F_i(x)|}\,,\]
and $\Theta(\eta) = \E\co\po\eta+G\pf_{+}\cf$ with $G \sim \mathcal{N}(0,1)$ a standard Gaussian variable.
The transition kernel $q_i(x,v,\dd v')$ is such that at the moment of a jump, the velocity is updated as
\begin{equation}\label{eq:majvitesse}
v \leftarrow v -\frac{2\veps(x)}{\sqrt{\beta}(1+\veps^2(x))}\po \eta + \widetilde{G} \pf \frac{M^{-1}F_i(x)}{|M^{-1/2}F_i(x)|},
\end{equation}
where $\widetilde{G}$ is a one-dimensional random variable with density  \[f_\eta(y) = \frac{1}{\Theta(\eta)\sqrt{2\pi}}(\eta+y)_{+} e^{-y^2/2}.\] In practice, depending on the context, the parameter $\veps$ will be chosen to be either a positive constant or proportional to the norm of the force: $\veps(x) = \veps_0|F_i(x)|$ with $\veps_0$ a positive constant. It is shown in \cite{MRZ} that in particular, $\veps = \infty$ corresponds to the BPS (i.e. the jump process described in~\eqref{eq:tauxbps} and~\eqref{eq:reflexion}), and that when $\veps \rightarrow 0$, the corresponding velocity jump process  (with a free transport on positions) converges to the Hamiltonian dynamics associated with the force $F_i$, i.e the process solving:
\[
\left\{\begin{array}{rcl}
\dd X_t & = &  V_t \dd t \\
\dd V_t & = & -M^{-1}F_i(X_t) \dd t\,. 
\end{array} \right.
\]
Moreover, as explained in SI, section 4, the definitions \eqref{eq:tauxsaut} and \eqref{eq:majvitesse} ensure that  the Gibbs measure $\mu$ is left invariant by the velocity jump Langevin process defined in Section~\ref{sec:proc-continu}, and the ergodicity of the process in a theoretical setting is proven in~\cite{splittinglangevinjumps}.
In other words, when $\veps$ is small enough, the velocity jump Langevin process is close, in terms of stochastic trajectories, to the Langevin dynamics, which yields (as we will see in Section~\ref{sec:simus}), a good preservation of dynamical properties of the system on top of a computational speedup without loss of accuracy in sampling.

\subsection{Splitting schemes and integration to multi-time-step methods}\label{sec:splitting}
\subsubsection{JUMP integrator}
In practice, since the solution of the SDE \eqref{eq:langevinbis} cannot be simulated exactly, time discretization methods need to be used to approximate the process. Similarly to the BAOAB~\cite{baoab2,stoltzsplitting,baoab1} scheme, the continuous-time dynamics of the velocity jump Langevin process can be approximated following a Trotter/Strang splitting scheme. We will describe three of them, and refer to the SI, section 2, for more technical details.

The first splitting consists in decomposing the dynamics into a free transport (A), an acceleration due to $F_0$ (B), a jump (J) and a friction-dissipation part (O). For a given time-step $\delta$, one step of the numerical scheme is given by the succession of steps BJAOAJB, where:
\begin{enumerate}
\item[(B)] $v \leftarrow v - \frac\delta2 M^{-1}F_0(x)$  \text{ (force $F_0$)}
\item[(J)] $v\leftarrow W_{\delta/2}$  \text{ (jumps)}
\item[(A)] $x \leftarrow x + \frac\delta2 v$  \text{ (free transport)}
\item[(O)] $v \leftarrow e^{-\gamma \delta} v + \sqrt{\beta ^{-1} \po 1 - e^{-2\gamma \delta  } \pf} M^{-1/2}G$ with $G\sim\mathcal{N}(0,I_d)$  \text{ (friction/dissipation)}\,,
\end{enumerate}
and where $(W_s)_{s\geqslant 0}$ is the piecewise-constant process initialized at $W_0=v$, that jumps at rate $\lambda_i$ according to $q_i$ for each $i\in\cco 1,K\ccf$. Such a scheme is similar to BAOAB, but where two half-time jump steps are added between the forces and the transport part. As explained and proved in~\cite{splittinglangevinjumps}, this palindromic form gives rise to a second-order scheme in the time-step, that is, the discretization bias on the invariant measure is of order $\delta^2$. This was the procedure used in~\cite{velocity} for the particular jump process defined in equations~\eqref{eq:tauxbps} and~\eqref{eq:reflexion}. For numerical stability reasons that we will detail in Section~\ref{sec:application}, it will be sometimes useful to merge the (B) and (J) parts into one (C) step, and run a splitting scheme of the form CAOAC, where the (C) corresponds to
\[ \text{(C) } v\leftarrow \widetilde{W}_{\delta/2}\,,\]
with $(\widetilde{W}_s)_{s\geqslant 0}$ is defined as $(W_s)_{s\geqslant 0}$, but such as between jumps, it undergoes the constant acceleration $-M^{-1}F_0(x)$ instead of being constant.

In fact, the BJAOAJB and CAOAC schemes can be combined: we can for instance treat some of the forces $F_i$ in (C), and the others separately in (J), that is, doing a scheme of the form C'J'AOAJ'C', where (J') treats, through a piecewise-constant velocity jump process, the forces $F_i$ for $i\in\cco 1,K'\ccf$ (with $K'<K$) and (C') treats the $F_i$ for $i\in\cco K'+1,K\ccf$ and $F_0$ in the same way as the step (C) described above. This situation will also be encountered in the following applications, see Section~\ref{sec:application}. We will now refer to those splitting schemes (BJAOAJB, CAOAC or C'J'AOAJ'C') as the JUMP integrator.

\subsubsection{JUMP-RESPA and JUMP-RESPA1 integrators}

All these steps can be integrated in a classical multi-timestep framework. Let us say that $F_0$ is itself a sum of two distinct terms $F_0 = F_{0,0}+F_{0,1}$, where, for instance, $F_{0,1}$ corresponds to a slow-varying many-body force, that cannot be treated efficiently with velocity jumps, but that doesn't need to be computed as often as the fast-varying forces gathered in $F_{0,0}$. In that case, let $Q$ be the transition kernel corresponding to a splitting scheme (like BJAOAJB, CAOAC or C'J'AOAJ'C') associated to $\nabla U - F_{0,1}$ with a certain time-step $\delta$. Let $\Delta = n\delta$ (with n$\in 2\N$) be a larger time step. The principle of the JUMP-RESPA scheme is to use the following splitting:
\begin{enumerate}
    \item $v\leftarrow v-\frac{\Delta}{2}M^{-1}F_{0,1}(x)$\,.
    \item Perform $n$ times the step $Q$ with time step $\delta$\,.
    \item $v\leftarrow v-\frac{\Delta}{2}M^{-1}F_{0,1}(x)$\,.
\end{enumerate}

This procedure can itself be iterated to include more different time steps, such as BAOAB-RESPA1~\cite{pushinglimits} that has three layers of time steps. By analogy, We will refer to this situation as the JUMP-RESPA1 integrator. More  details are given in SI, section 2.

Let us now see how to simulate efficiently the processes $(W_s)_{s\geqslant 0}$ and $(\widetilde{W}_s)_{s\geqslant 0}$, and how this gives rise to a computational advantage.

\subsection{Efficient simulation of jumps}\label{sec:thinning}
Let us start with the process $(\widetilde{W}_s)_{s\geqslant 0}$ (corresponding to the (C) step in the CAOAC scheme). We suppose that each jump rate $\lambda_i$ is bounded by a constant $\overline{\lambda_i}$, and denote $\overline{\lambda} = \sum_i \overline{\lambda_i}$. Then an easy computation on the generator (see SI, section 3) shows that the jumps can be simulated  exactly this way: starting from $(x,v)$,
\begin{enumerate}
    \item Draw $\mathcal{E}$ a standard exponential random variable, and let $T = \mathcal{E}/\overline{\lambda}$ be the next jump time proposal.
    \item Draw $I$ such that $\PP (I=i) = \overline{\lambda_i}/\overline{\lambda}$. Propose a jump of type $I$ at time $T$.
    \item The jump is accepted with probability $\lambda_I(x,v-M^{-1}F_0(x)T)/\overline{\lambda_I}$, in which case the velocity is resampled at time $T$ according to $q_I(x,v - M^{-1}F_0(x)T,\dd v')$, otherwise the velocity at time $T$ is simply $v - M^{-1}F_0(x)T$ (there is no jump).
\end{enumerate}
The construction is then repeated by induction over the jump time proposals. The case of $(W_s)_{s\geqslant 0}$ is very similar: the only difference is that since the process is constant between jumps, in the step 3, a jump of type $I$ is accepted with probability $\lambda_I(x,v)/\overline{\lambda_I}$, in which case the velocity is resampled according to $q_I(x,v,\dd v')$. This method, called Poisson thinning \cite{thinning1,thinning2} (which is exact, in the sense that it does not induce any further time discretization error) is the key to the computational advantage of the algorithm: contrary to a classical numerical scheme of the Langevin diffusion (such as BAOAB) where the gradient $\nabla U$ is computed at each time step, here we only need to evaluate $F_i$ when a jump of type $i$ is proposed (at the step 3 above), which does not happen at each time step for each $i$. However, notice that similarly to BAOAB and contrary to multi-timestep splitting methods, there is still a unique time step in the discretization and no additional parameters to tune: the frequency at which $F_i$ is evaluated is random and is adapted to each force.

In many cases, we will use the same bound for all the forces $F_i$, so the $\overline{\lambda_i}$ terms will be uniform in $i$ : $\overline{\lambda_i} = \lambda^{*}$ (so that $\overline{\lambda} = K\lambda^*$). In those cases, the step $2$ above is simply a uniform draw among all the $i$. The Algorithm 1 describes the (C) step of the CAOAC scheme, and the Algorithm 2 describes the (J) step of the BJAOAJB scheme.

In fact, as shown in SI, section 6, in the general case, the bounds $\overline{\lambda}_i$ are local (they depend the velocity $v$) and need to be re-evaluated after each jump. This is not a problem, since the only requirement in order to have a computational advantage is that those bounds do not depend on $F_i$. However, in the case of the Bouncy Particle Sampler (i.e. the jump mechanism defined in~\eqref{eq:tauxbps} and~\eqref{eq:reflexion}), the norm of the velocity doesn't change after a jump, so $\overline{\lambda}$ stays constant throughout the time step. In those cases, when we simulate the process up to a time $\delta$, we don't need to know exactly the jump times, but only the number of jumps and the order in which they occur.
Classical properties of the exponential distribution ensure that during the time interval $\co 0, \delta \cf $, the number of jump proposals (step 1 above) follows a Poisson distribution of parameter $\overline{\lambda}\delta$. Then the simulation procedure for the (J) part of BJAOAJB is the following: starting from $(x,v)$,

\begin{enumerate}
    \item Draw $M \sim \text{Poisson}(\overline{\lambda}\delta)$ (the total number of jump proposals). 
    \item For each $j\in\cco 1, M \ccf$
        \begin{enumerate}
            \item Draw $I$ such that $\PP (I=i) = \overline{\lambda_i}/\overline{\lambda}$. The $j$-th proposal is a jump of type $I$.
            \item Accept the proposal (i.e. resample the velocity according to $q_I(x,v,\dd v')$) with probability $\lambda_I(x,v)/\overline{\lambda_I}$.
        \end{enumerate}
\end{enumerate}
This situation is described in Algorithm 3.

\begin{algorithm}[htbp]
\caption{Simulation of (C) part.\newline
Simulates the jump process $(\widetilde{W}_s)_{s\geqslant 0}$ for a time $\delta/2$, starting from $(x,v)$.
}
\begin{algorithmic}[1]
\Statex
\Procedure{Velocity-update}{$\delta,x,v$}
    \Statex    
    \State $T\gets 0$, $t\gets 0$ \Comment{$t$ is the current time, $T$ is the time of jump proposals}
        \State $a \gets -M^{-1}F_0(x)$ \Comment{a is the acceleration due to the short range forces}
        \While{$t\leqslant \delta/2$}
            \State $T \gets \mathcal{E}(K\lambda^*)$ \Comment{Draw the next jump proposal time: an exponential law of parameter $\overline{\lambda}$}
            \If{$T+t > \delta/2$}
                \State $v \gets v+(\delta/2-t)a$
            \Else
                \State $t \gets t+T$ \Comment{Update the current time}
                \State $v \gets v + Ta$ \Comment{Update the velocity until the jump proposal}
                \State $I \gets$ \Call{Random}{$\cco 1,K \ccf$} \Comment{Choose a jump type}
                \State $U \gets$ \Call{Random}{$[0,1]$}
                \State $\lambda \gets \lambda_I(x,v)$ \Comment{compute the  I-th jump rate at the current state} 
                \If{$U\leqslant \lambda/\lambda^*$}
                    \State $v \gets$ \Call{Sampling}{$q_I(x,v,\dd v')$} \Comment{the velocity is resampled}
                \EndIf
            \EndIf
        \EndWhile
\EndProcedure
\end{algorithmic}
\end{algorithm}

\begin{algorithm}[htbp]
\caption{Simulation of (J) part (general case).\newline
Simulates the piecewise-constant jump process $(W_s)_{s\geqslant 0}$ for a time $\delta/2$, starting from $(x,v)$.
}
\begin{algorithmic}[1]
\Statex
\Procedure{Jump}{$\delta,x,v$}
    \Statex    
    \State $T\gets 0$, $t\gets 0$ \Comment{$t$ is the current time, $T$ is the time of jump proposals}
        \While{$t\leqslant \delta/2$}
            \State $T \gets \mathcal{E}(K\lambda^*)$ \Comment{Draw the next jump proposal time: an exponential law of parameter $\overline{\lambda}$}
            \If{$T+t > \delta/2$}
                \State $t \gets \delta/2$
            \Else
                \State $t \gets t+T$ \Comment{Update the current time}
                \State $I \gets$ \Call{Random}{$\cco 1,K \ccf$} \Comment{Choose a jump type}
                \State $U \gets$ \Call{Random}{$[0,1]$}
                \State $\lambda \gets \lambda_I(x,v)$ \Comment{compute the  I-th jump rate at the current state} 
                \If{$U\leqslant \lambda/\lambda^*$}
                    \State $v \gets$ \Call{Sampling}{$q_I(x,v,\dd v')$} \Comment{the velocity is resampled}
                \EndIf
            \EndIf
        \EndWhile
\EndProcedure
\end{algorithmic}
\end{algorithm}

\begin{algorithm}[htpb]
\caption{Simulation of (J) part (constant bound case)\newline
Special case: a jump does not affect the bound on the jump rate. \newline
Simulates the piecewise-constant jump process $(W_s)_{s\geqslant 0}$ for a time $\delta/2$, starting from $(x,v)$.
}
\begin{algorithmic}[1]
\Statex
\Procedure{Jump}{$\delta,x,v$}
    \Statex
    \State $M \gets$ \Call{Poisson}{$K\lambda^*\delta/2$}
    \For{$j \in \cco 1,M\ccf$}
        \State $I \gets$ \Call{Random}{$\cco 1,K\ccf$} \Comment{For each jump proposal $j$, choose randomly a jump type $I$}
        \State $U \gets$ \Call{Random}{$[0,1]$}
        \State $\lambda \gets \lambda_I(x,v)$ \Comment{compute the  I-th jump rate at state $(x,v)$}
        \If{$U \leqslant \lambda/\lambda^*$}
            \State $v \gets$ \Call{Sampling}{$q_I(x,v,\dd v')$} \Comment{the velocity is resampled}
        \EndIf
    \EndFor
    \State \textbf{return}
    \Statex
\EndProcedure
\end{algorithmic}
\end{algorithm}

Regarding the simulation of the jump itself, it is shown in \cite{MRZ} that the random variable $\widetilde{G}$ that appears in~\eqref{eq:majvitesse} 
can be easily simulated with rejection sampling. Assuming that we can sample a distribution with density $g$ that satisfies $f(x) \leqslant C g(x)$ and such that the ratio $\frac{f(x)}{C g(x)}$ is easy to compute for any $x$, the procedure is the following: 
\begin{enumerate}
    \item Draw a proposal $Y$ with density $g$.
    \item Draw $U$ uniform in $[0,1]$. if $U\leqslant \frac{f(Y)}{C g(Y)}$ then the proposal is accepted.
    \item Otherwise, repeat from step 1.
\end{enumerate}
The mean number of proposals before acceptation is $C$, so the smaller the constant the faster the procedure is. Hence, following  \cite{MRZ} (to which we refer for details), in our case,  $g$ is a:
\[\left\{\begin{array}{rcl}
&\text {Gamma law }& \text{for } \eta  < -2.5 \\
&\text {Exponential law}&\text{for }   -2.5 \leqslant \eta < -1 \\
&\text {Rayleigh law }& \text{for } -1 \leqslant  \eta \leqslant 0 \\
&\text {Mixed Rayleigh-Gaussian law}& \text{for }   \eta > 0.
\end{array} \right.\]

\section{Application to electrostatic and van der Waals interactions}\label{sec:application}

In a classical (non-polarizable) force field such as OPLS \cite{opls}, AMBER \cite{wang2000well} or CHARMM \cite{CharmmFF}, the non-bonded parts of the interaction potential are the Lennard-Jones and electrostatic contributions. Since the Lennard-Jones term decreases very fast, in periodic boundary conditions one can simply compute the forces in a radius up to a certain cutoff. The electrostatic part, decreasing at a much slower rate, is more problematic. A classical way of treating the long range interactions is by using Ewald summation techniques, for instance through the SPME method \cite{spme} (although other approaches exist, see for instance~\cite{RESPA1switch}). In this method, the Coulomb potential is decomposed in a direct and a reciprocal part, the latter being computed in the Fourier space with a fast Fourier transform. In practice, the algorithm BOUNCE described in \cite{velocity}, i.e. a BJAOAJB splitting on the process with the BPS jump mechanism, has been implemented in Tinker-HP \cite{tinkerhp} and tested in periodic boundary conditions with jumps on the long-range parts of the van der Waals interactions and the direct part of the electrostatic interactions. The reciprocal part was treated with a classical multi-timestep method, the reference system propagator algorithm (RESPA)\cite{TuckermanRossiBerne}. Since electrostatics are much stronger than van der Waals at a long range, the BPS electrostatic jumps can only be done on a very small range. Jumping on a larger range yields too many velocity bounces, in the sense that the numerical scheme with the classical timestep of $1fs$ is very unstable. One solution would be to decrease the timestep, which is of course counterproductive, since the goal is to reduce the final cost of the calculations.

On the contrary, the jump mechanism defined in~\eqref{eq:tauxsaut} and~\eqref{eq:majvitesse} yields milder jumps than the othogonal reflexions of the BPS. Moreover, as discussed earlier, the process can be seen as an interpolation between bounces and diffusions: in the situations where BOUNCE is numerically unstable, choosing a small parameter $\veps$ makes the process closer to classical Hamiltonian dynamics. In addition to that, combining the $(B)$ and $(J)$ parts (i.e. doing the CAOAC scheme instead of BJAOAJB) stabilizes further the discretized process.

These two elements combined, namely a more general jump process and the fusion of the ``jump" and ``force" steps, allow to replace a larger range of the potential by jumps than in the BOUNCE algorithm, yielding more computational gain, while preserving dynamical properties of the system such as the diffusion coefficients. Finally, the BPS jumps of \cite{velocity} on the van der Waals forces can also be integrated in the general procedure, and two multi-timestep versions of the algorithm, namely JUMP-RESPA and JUMP-RESPA1 combine all type of jumps and yields a computational gain over the BAOAB-RESPA and BAOAB-RESPA1 \cite{pushinglimits} algorithm.

\subsection{Direct electrostatic jumps}\label{sec:direct}

As reminded in the SI, section, 5, the Ewald summation splits 
the electrostatic interactions into a direct and a reciprocal part. In Periodic Boundary Conditions (PBC), this decomposition can be chosen in a way that enforces the ``minimum image convention": in the direct space, each atom interacts only with the closest image of an other atom. We assume in the following that this condition is satisfied. By denoting $r_{ij}=|x_i-x_j|$ the distance between the atoms $i$ and $j$, the direct electrostatic energy associated to each pair $(i,j)$ is:
\[U_{ij}(x) = \frac{q_i q_j}{r_{ij}}\text{erfc}(\alpha r_{ij}).\]
In order to decompose this term into a short and a long range contribution, let $r_c$ be a cutoff (smaller than the ``Ewald cutoff" that separates the direct and reciprocal part, whose value is enforced by the choice of the parameter $\alpha$), $\chi(r)$ a switching function that goes smoothly from 1 to 0 around the cutoff (its precise formula is given in SI, section 6) and let
\[U_{ij} = \chi(r_{ij})U_{ij} + (1-\chi(r_{ij}))U_{ij}\,.\]
We now denote $F_{ij} = \nabla_{x_i}(1-\chi(r_{ij}))U_{ij}\in\R^{3}$, the component of the long-range force of the atom $j$ acting on the atom $i$, Therefore, the direct part of electrostatic interactions is decomposed into
\[\nabla U_{direct} = F_0 + \sum_{i\neq j} A_iF_{ij}\,,\]
where $A_i\in\mathcal{M}_{3N\times 3}$ is the matrix with all coefficients are equal to zero except $A(3i-2,1)=A(3i-1,2)=A(3i,3)=1$. Therefore, a jump mechanism is associated to each pair interaction $A_{i}F_{ij}$. In this context,~\eqref{eq:tauxsaut} and~\eqref{eq:majvitesse} have simpler expressions than in the general case described in Section~\ref{sec:nummethod}. Here, $|M^{-1/2}F_{ij}(x)| = \frac{|F_{ij}(x)|}{\sqrt{m_i}}$ (with $m_i$ the mass of the atom $i$), which yields the following jump rate:
\[ \lambda_{ij}(x,v) = \frac{\sqrt{\beta}}{\veps(x)\sqrt{m_i}}|F_{ij}(x)|\Theta\po\veps(x)\sqrt{\beta m_i}\frac{v_i\cdot F_{ij}(x)}{|F_{ij}(x)|}\pf\,,\]
and a jump only modifies the atom $i$:
\[v_i \leftarrow v_i - \frac{2\veps(x)}{1+\veps^2(x)}\po \veps(x)\frac{v_i\cdot F_{ij}(x)}{|F_{ij}(x)|} + \frac{\widetilde{G}}{\sqrt{\beta m_i}} \pf \frac{F_{ij}(x)}{|F_{ij}(x)|},\]
where $\widetilde{G}$ has been generated from the distribution $f_\eta$ with $\eta = \veps(x) \sqrt{\beta m_i}\frac{v_i\cdot F_{ij}(x)}{|F_{ij}(x)|}$. 
As detailed in the SI, section 6, in this case it is easy to obtain explicit analytic bounds on $|F_{ij}|$ that are required in the thinning method. Moreover, for each $i$, those bounds are uniform in $j$ which, as explained in Section~\ref{sec:thinning}, simplifies the thinning procedure.

Since the jump mechanisms associated with $F_{ij}$ only involve the $i$-th atom, the jump processes associated to the different atoms can be seen as independent Markov chains that can be simulated in parallel, which in particular allows massively parallel implementations on GPU, see Section~\ref{sec:simus}.

Finally, as mentioned in the introduction of this section, in this case it is more convenient to perform a CAOAC scheme instead of BJAOAJB. If we treat a large portion of the interactions with jumps, the rate $\lambda_{ij}$ will be higher, yielding many velocity jumps, so merging the (J) and (B) steps of the splitting scheme has the effect of stabilizing the dynamics.
The Algorithm 4 below describes precisely the procedure when $\veps$ is constant. The case of an adaptive $\veps(x) = \veps_0|F_i(x)|$ is exactly the same but with a slightly different expression for the jump rate, the bound and the kernel.

\begin{algorithm}[htpb]
\caption{Step (C): a general velocity update (with constant $\veps$)\newline
Simulates during a time $\delta/2$ the PDMP treating the acceleration due to the short range forces and the jumps due to the long range forces. The position $x$ is fixed.\\
\textbf{Input: } $N$ the number of atoms, $(x,v)$ their positions and velocities, $\delta$ the time step
}
\begin{algorithmic}[1]
\Statex
\Procedure{VelocityUpdate}{$N,\delta,x,v$}
    \Statex
    \For{$i \in \cco 1,N\ccf$}
        \State $T\gets 0$, $t\gets 0$ \Comment{$t$ is the current time, $T$ is the time of jump proposals}
        \State $v_i \gets v_i^0$ \Comment{$v_i^0$ is the initial velocity of atom $i$}
        \State $a_i \gets -\nabla U_{\text{short}}(x)/m_i$ \Comment{a is the acceleration due to the short range forces}
        \State $L \gets L_i$ \Comment{The uniform bound on the $|F_{ij}|, 1\leq j\leq N$}
        \While{$t\leqslant \delta/2$}
            \State $B_v \gets (|v_i|^2+(\delta/2-t)^2|a_i|^2+2(\delta/2-t)(v_i\cdot a_i)_{+})^{1/2}$ \Comment{bound on $|v_i|$ during the time $[t,\delta/2]$}
            \State $\overline{\lambda} \gets L\times N\po\beta B_v + \frac{1}{\veps}\sqrt{\frac{\beta}{2\pi m_i}}\pf $ \Comment{The bound on the jumprate}
            \State $T \gets \mathcal{E}(\overline{\lambda})$ \Comment{Draw an exponential law of parameter $\overline{\lambda}$}
            \If{$T+t > \delta/2$}
                \State $v_i \gets v_i+(\delta/2-t)a_i$
            \Else
                \State $t \gets t+T$
                \State $v_i \gets v_i + Ta_i$
                \State $\eta \gets \veps \sqrt{\beta m_i}\frac{v\cdot F_{ij}(x)}{|F_{ij}(x)|} $
                \State $J \gets$ \Call{Random}{$\cco 1,N \ccf$}
                \State $U \gets$ \Call{Random}{$[0,1]$}
                \State $p \gets \lambda_{ij}(x,v)/\overline{\lambda} = \frac{|F_{ij}(x)|\Theta(\eta)}{L\po B_v\veps\sqrt{m_i}+1/\sqrt{2\pi}\pf}$\Comment{the probability of acceptation}
                \If{$U\leqslant p$}
                    \State $\widetilde{G} \gets $ \Call{RejectionSampling}{$\eta$} \Comment{Generate $\widetilde{G}$}
                    \State $v_i \gets v_i - \frac{2\veps}{\sqrt{\beta m_i}(1+\veps^2)}\po \eta +\widetilde{G} \pf \frac{F_{i,j}(x)}{|F_{i,j}(x)|}$ \Comment{Performs a jump}
                \EndIf
            \EndIf
        \EndWhile
    \EndFor
    \State \textbf{return}
    \Statex
\EndProcedure
\end{algorithmic}
\end{algorithm}

\subsubsection{Optimization of the bound: the ring technique}

Let $r_E$ be the Ewald cutoff, i.e. the maximum radius that is taken into account in the direct part and $r_c$ the cutoff between the short-range interactions treated in the acceleration part and the long-range interactions treated with the jumps. In order to improve the bound on the jump rate (and therefore reduce the number of jump proposals), we can introduce an intermediate cutoff $r_M$ ($r_c<r_M<r_E$) and compute two different bounds: a bound on the $|F_{ij}|$ such that $r_c-h \leq r_{ij} \leq r_M $ ($h$ is the switching parameter) denoted $L_M$ and a bound on the $|F_{ij}|$ such that $r_M \leq r_{ij} \leq r_E $ denoted $L_L$. For each $i$, we denote $N_M(i)$ and $N_L(i)$ the number of atoms $j$ in these two cases respectively. Now, the term $N\times L_i$ in the bound on the jump rate $\overline{\lambda_i}$ can be replaced by $N_M(i)L_M+N_L(i)L_L$. The algorithm (again in the case where $\veps$ is constant) can then be modified the following way:
\begin{enumerate}
    \item Draw a proposition at rate $\overline{\lambda_i}$.
    \item Draw $Y\in\{M,L\}$ such that $\PP(Y=M) = \frac{L_MN_M(i)}{N_M(i)L_M+N_L(i)L_L}$ and $\PP(Y=L) = \frac{L_LN_L(i)}{N_M(i)L_M+N_L(i)L_L}$.
    \item Draw uniformly $j$ in the right neighbor-lists (namely the ``middle-range" or the ``long-range" list).
    \item Accept a jump of type $(i,j)$ with probability $\frac{|F_{ij}(x)|\Theta(\eta)}{L_Y\po B_v\veps\sqrt{m_i}+1/\sqrt{2\pi}\pf}$
\end{enumerate}

\begin{remark}
    This modification requires to build new ``middle" and ``long" neighbor-lists, in order to know $N_L(i)$ and $N_M(i)$ for each atom and to draw uniformly among them.
\end{remark}

\subsection{Lennard-Jones jumps}\label{sec:jumpsvdw}

The same procedure can be used for van der Waals forces. In this case, the potential is given by:
\[U_{\text{vdW}}(x_1,\dots,x_N) = \sum_{i\neq j}\veps_{ij}\co\po\frac{\sigma_{ij}}{r_{ij}}\pf^{12} - \po\frac{\sigma_{ij}}{r_{ij}}\pf^6\cf.\]
for some parameters $\sigma_{ij},\veps_{ij}>0$. Since each term of this sum decreases very fast in $r_{ij}$, there is no splitting in direct and reciprocal parts and all the terms beyond a certain cutoff $r_{\text{vdW}}$ (i.e. for $r_{ij}>r_{\text{vdW}}$) are neglected.
Again, we can define the pairwise term
\[U_{ij}^{\text{vdW}}(x) = \veps_{ij}\co\po\frac{\sigma_{ij}}{r_{ij}}\pf^{12} - \po\frac{\sigma_{ij}}{r_{ij}}\pf^6\cf, \]
choose a cutoff $r_c<r_{\text{vdW}}$ and use the switching function $\chi$ to separate each term into a short-range and a long-range part:
\[U_{ij}^{\text{vdW}} = \chi(r_{ij})U_{ij}^{\text{vdW}} + (1-\chi(r_{ij}))U_{ij}^{\text{vdW}}\,.\]
From now, we define $F_{ij} = \nabla_{x_i}(1-\chi(r_{ij}))U_{ij}$ and build a jump mechanism associated to each couple $(i,j)$. Here, since the van der Waals forces are weaker than the Coulomb ones at long range, there are less stability issues and we can choose the parameter giving the least jumps: $\veps = \infty$, which is the case that corresponds to the BPS. The jump rate associated to the force $F_{ij}$ is:
\[\lambda_{ij}(x,v) = \beta\po v_i\cdot F_{ij}(x)\pf_{+}\,,\]
and the jumps are deterministic: $q_{ij}(x,v,\dd v')=\delta_{R_{ij}(x,v)}(\dd v')$ where
\[R_{ij}(x,v) = v - 2 \frac{v\cdot F_{ij}(x)}{|F_{ij}(x)|^2}F_{ij}(x).\]
If we denote $L_i$ a bound on all the $|F_{ij}|$ for $1\leqslant j \leqslant N$, the jump rate can be bounded by:
\[\lambda_{ij}(x,v) \leqslant \beta|v|L_i := \lambda_{ij}^*\,.\]
A jump being an orthogonal reflexion of $v$ against $F_{ij}(x)^\perp$, it does not modify $|v|$ and the bound $\lambda_{ij}^*$ can stay the same during all the time step. It is therefore advantageous to use a BJAOAJB splitting with the Algorithm~3 described in \ref{sec:thinning}. Moreover, as in the case of jumps on the direct electrostatic part, $\lambda_{ij}^*$ is the same for all $j$, we can define $\overline{\lambda_i} = N\lambda_{ij}^*$ and the process associated to the velocity of each atom can be simulated independently. This yields the Algorithm 5, that was also described in \cite{velocity}.

\begin{algorithm}[htpb]
\caption{Step (J): BPS for van de Waals interactions\newline
Simulates the piecewise-constant jump process for a time $\delta/2$, starting from $(x,v)$.\newline
\textbf{Input :} $N$ the number of atoms, $(x,v)$ their positions and velocities, $\delta$ the time step
}
\begin{algorithmic}[1]
\Statex
\Procedure{Jump}{$N,\delta,x,v$}
    \Statex
    \For{$i \in \cco 1, N\ccf$}
    \State $M_i \gets$ \Call{Poisson}{$N\beta|v_i|L_i\delta/2$} \Comment{Draw the number of jump proposals}
    \For{$k \in \cco 1,M_i\ccf$}
        \State $J \gets$ \Call{Random}{$\cco 1,N\ccf$} \Comment{For each jump proposal $k$, choose randomly a jump of type $J$}
        \State $U \gets$ \Call{Random}{$[0,1]$}
        \State $p \gets \frac{\po v_i\cdot F_{iJ}(x)\pf_+}{|v_i|L_i}$ \Comment{compute the probability of accepting the jump}
        \If{$U \leqslant p$}
            \State $v_i \gets R_{ij}(x,v)$ \Comment{the velocity bounces}
        \EndIf
    \EndFor
    \EndFor
    \State \textbf{return}
    \Statex
\EndProcedure
\end{algorithmic}
\end{algorithm}

\subsection{Conclusion: three new integrators}
To conclude, using the jump mechanisms described in this section gives rise to three new integrators.
First, the JUMP integrator consists in running either a BJAOAJB scheme, where the (J) part treats the Lennard-Jones jumps, a CAOAC scheme, where the (C) part treats the direct electrostatic jumps (along with the acceleration due to the rest of the forces) or, combining both, a C'J'AOAJ'C' scheme, where (J') treats the Lennard-Jones jumps, and (C') the electrostatic direct jumps.
Regarding the multi-timesteps versions, JUMP-RESPA simply consists in treating in a small time-step $\delta$ all the intramolecular forces of the potential (bond stretching, angle bending, bond-angle cross terms, out-of-plane bending and torsional rotations), and leaving all the non-bonded forces (namely the van der Waals and Coulomb interactions) in a larger time step $\Delta = n\delta$ (with $n\in 2\N)$, including the jumps. However, when using jumps, this is not very natural, as it increases the issues related to time discretization (unstability or numerical bias) while having no effect on the complexity of the jump parts, since with the thinning method, a step-size $n\delta$ requires on average $n$ times more jumps (hence computations) than a step-size $\delta$.

This remark leads to the following scheme:
in JUMP-RESPA1, a small time step $\delta$ treats all the intramolecular forces, an intermediate time step $\kappa = m\delta$ treats the short-range van der Waals, the short-range direct electrostatics and the jumps (that take into account the long range van der Waals forces and the long range direct electrostatics), and a large time step $\Delta=n\kappa$ treats the reciprocal electrostatics, the many-body force that is, in practice (especially in large systems) the most numerically expensive force to compute. If we applied the algorithm to a polarizable force field such as AMOEBA~\cite{amoeba}, the expensive polarization part would also be treated in this largest time step.

In comparison, in BAOAB-RESPA1, the long range van der Waals and long range direct space electrostatics are treated in the larger time step. Here, in JUMP-RESPA1, those are treated by the jumps, in the intermediate time step. As discussed above, indeed, the jump process has the property of not requiring more computations when put in a smaller time step: the jump rate stays exactly the same.

\section{Numerical experiments}\label{sec:simus}
All variations of the algorithms described in Section~\ref{sec:application} have been implemented in Tinker-HP software for molecular dynamics, both on the CPU and the parallel GPU versions. The software will be made freely available to academic users within the next release of the Tinker-HP code~\cite{githubtinkerhp,SiteTinkerHP,tinkerhpgpu,tinkerhp}. Computations were done on the computing cluster of the Theoretical Chemistry Laboratory (LCT) of Sorbonne University, Paris, on a 22 cores Intel(R) Xeon(R) Silver 4116 CPU @ 2.10GHz for the CPU version, and a Nvidia Quadro GV100 for the GPU version.
All numerical tests were done on water systems in the NVT ensemble with temperature  $T=300K$, of sizes ranging from 216 to 96000 molecules, with SPC/Fw model in the OPLS-AA force-field, in periodic boundary conditions. Simulation parameters are given at the end of this section, and SI, section 7, gives more details on the implementation.

\paragraph{Tuning of parameters.} As explained in Section~\ref{sec:thinning}, treating certain forces with jumps allows to reduce the total number of computations, and therefore accelerates the simulation. However, replacing a too large range of interactions by jumps yields either numerical instabilities (as it is the case for the BOUNCE algorithm described in~\cite{velocity}) or a significant loss in the auto-diffusion constant, which indicates a reduction of the sampling rate. As mentioned in \cite{pushinglimits} the ideal situation is when the acceleration rate is higher than the loss rate in the diffusion constant.

In order to do that, the choice of an adaptive $\veps$ helps the conservation of the diffusion, as the jump process will yield milder jumps in the shorter-range regions (where the forces are stronger). Moreover, the section of pairwise interaction that are replaced by jumps should not be too large, or in too short-range regions. In other words, there should not be too many jumps, either because of too many jump processes running in parallel (if too many interactions are treated by jumps) or because the jump rates are too high (if those interactions are too strong). Those settings, although they can push further the computational speedup, always yields a loss in auto-diffusion.

In other words, the best choice is to always choose an adaptive $\veps$ parameter, with a small enough $\veps_0$ (so that the trajectories stay close to the Langevin process), but not too small so there is still a computational advantage, and to replace either a small portion of relatively short-range electrostatic interactions (for example between $5$ and $7$ Angstroms), or a larger portion of longer-range interactions (for example between $7$ and $11$ Angstroms).

The first setting seems to be the best for small systems, and the second one is more appropriate for larger systems, especially on the GPU version. Indeed, treating very long range interactions with jumps allows to reduce the reciprocal Coulomb force, which is very expensive in large systems (it can take up to $50\%$ of the total computation time). Moreover, since the reciprocal Coulomb force is partly responsible for resonance issues in multi-timestep integrators, reducing this part allows to extend the stability limit of the external time step for the JUMP-RESPA1 algorithm, which allows for a further increase in simulation speed.

\begin{figure}[htbp]
\begin{center}
    \subfloat[Oxygen-Oxygen radial distributions]{\includegraphics[width=0.5\textwidth]{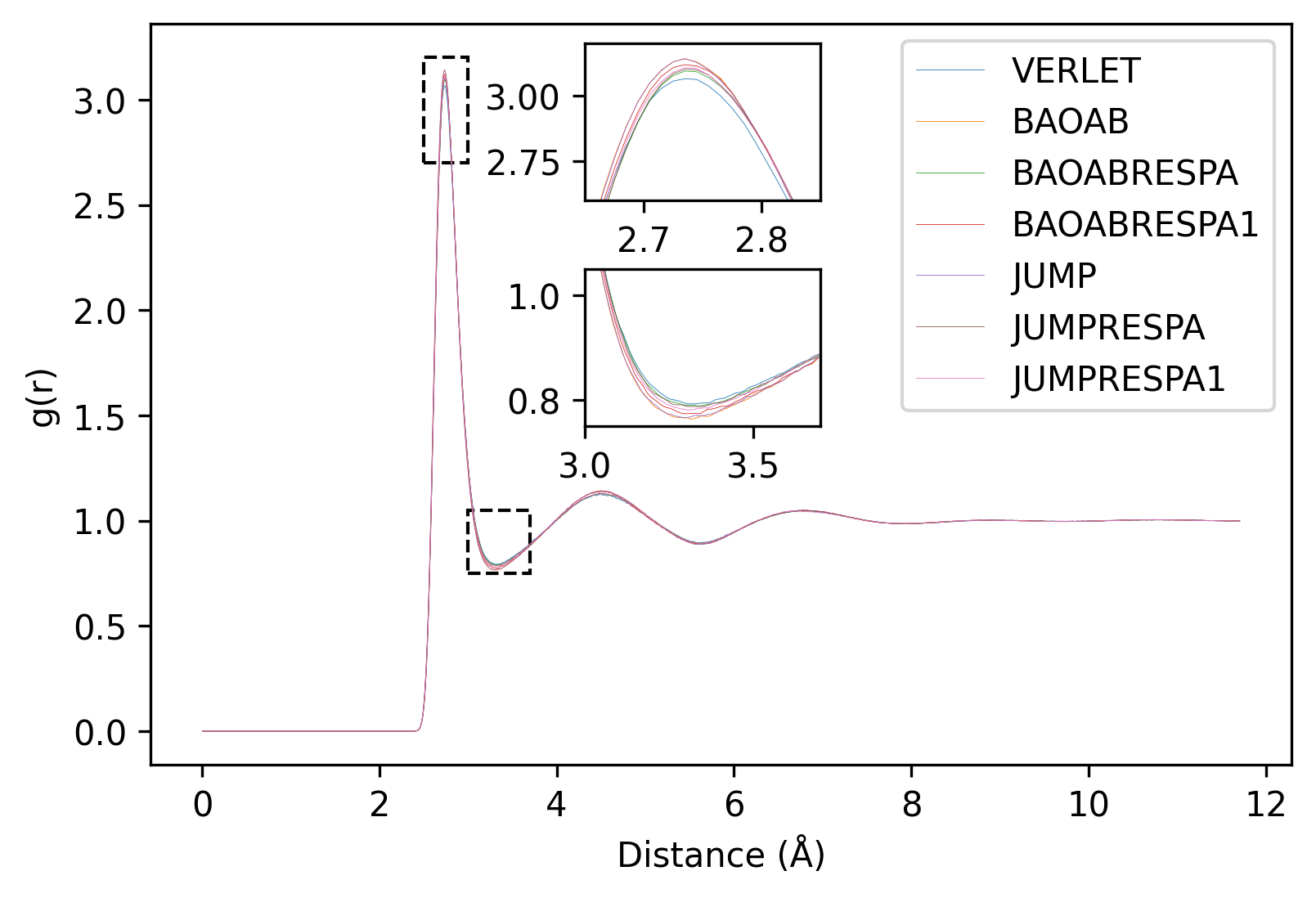}}
    \subfloat[Oxygen-Oxygen radial distributions without Coulomb forces between $5$ and $9$ Angström]{\includegraphics[width=0.5\textwidth]{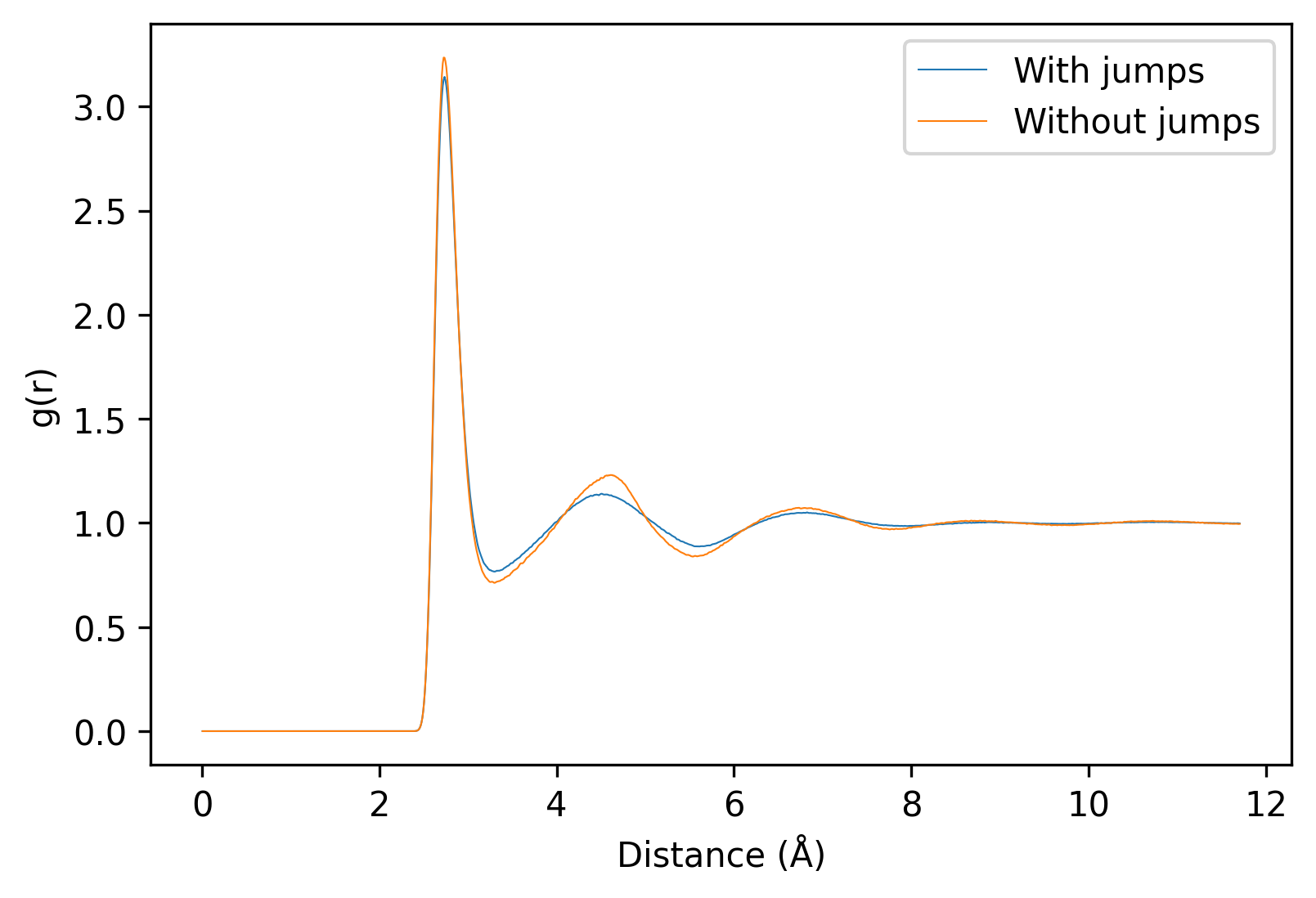}} \\
    \subfloat[Mean temperature]{\includegraphics[width=0.5\textwidth]{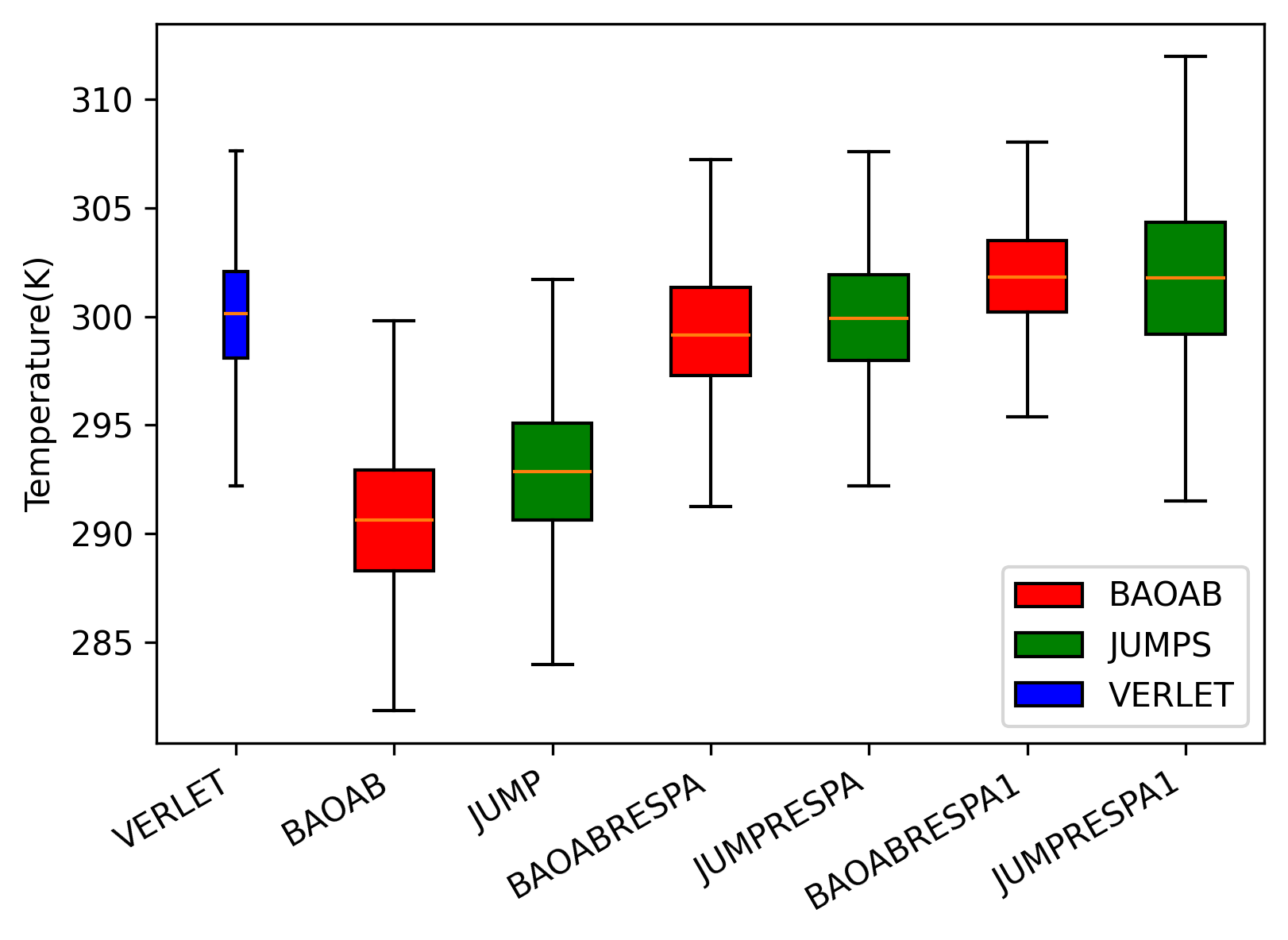}}
    \subfloat[Mean potential energy]{\includegraphics[width=0.5\textwidth]{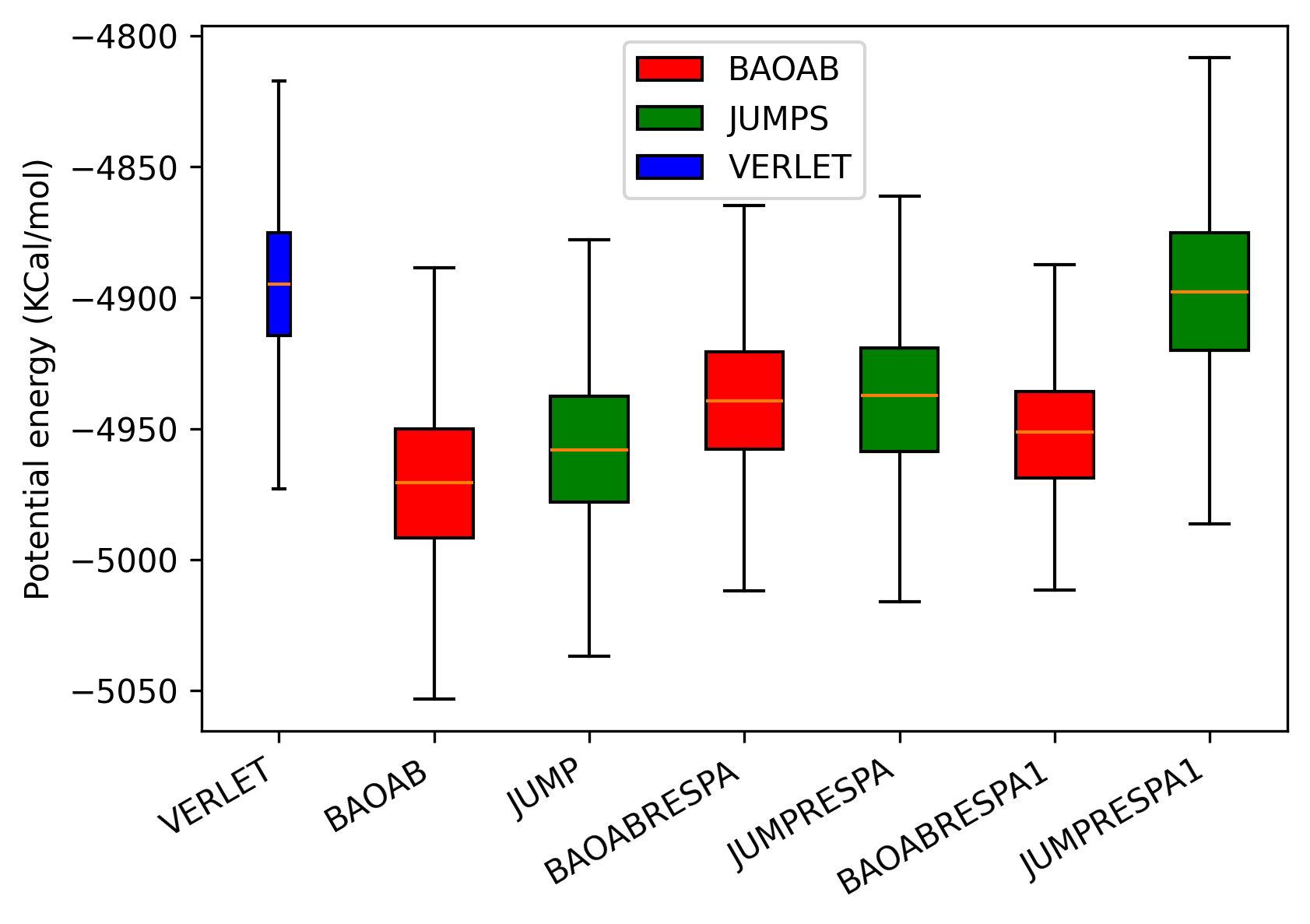}}    
\end{center}
\caption{Sampling properties of the JUMP algorithms for a system of 500 water molecules.}\label{fig:1}
\end{figure}

\begin{figure}
    \centering
    \includegraphics[width=0.8\linewidth]{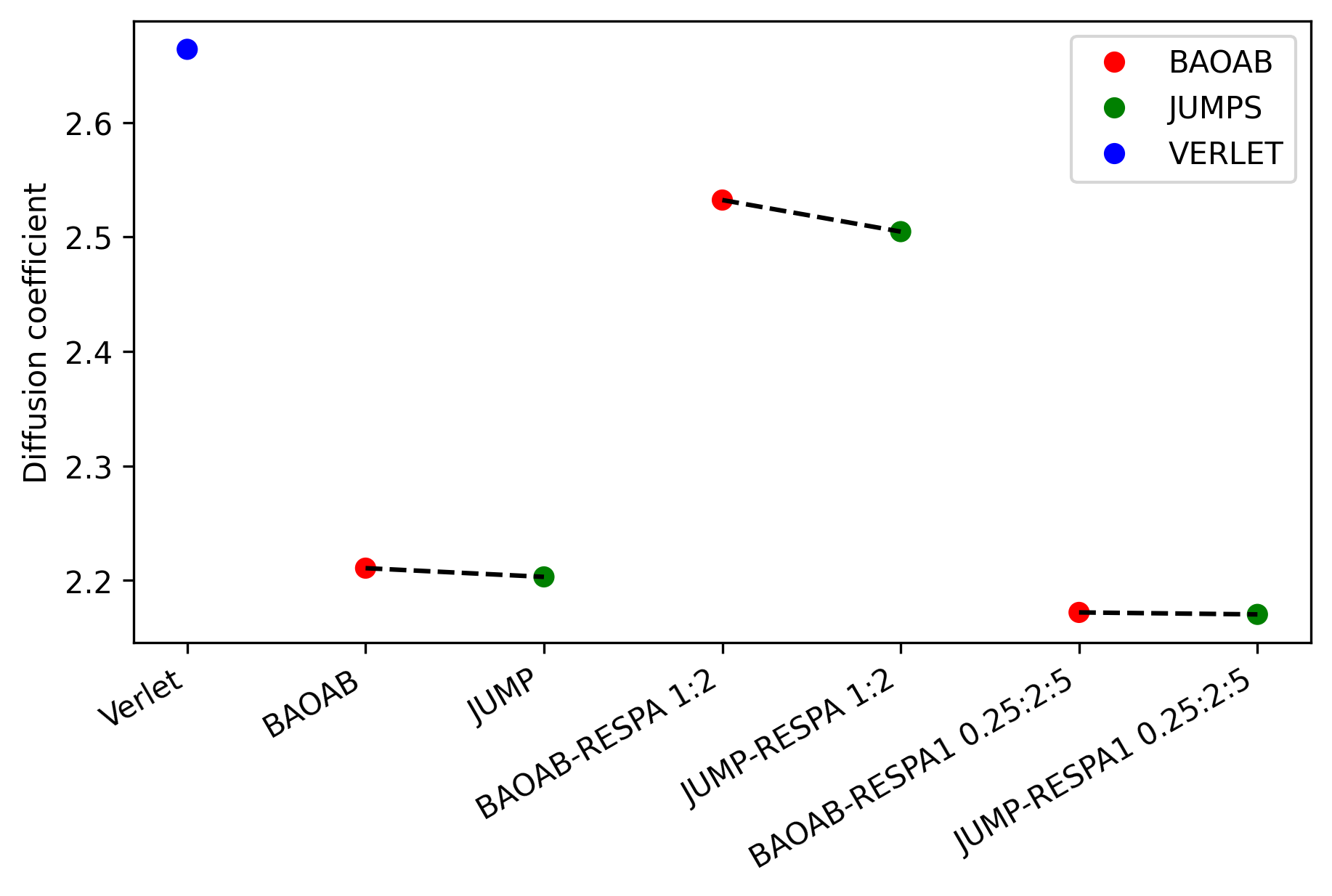}
    \caption{Diffusion coefficients for a system of 500 water molecules.}
    \label{fig:2}
\end{figure}

\paragraph{Precision in sampling and dynamical properties.} Figure~\ref{fig:1} shows the oxygen-oxygen radial distributions, the mean temperature and mean potential energy of $500$ water molecules, comparing the new JUMP, JUMP-RESPA and JUMP-RESPA1 integrators to its classical BAOAB, BAOAB-RESPA and BAOAB-RESPA1 counterparts, as well as the VERLET integrator for reference, with $2ns$ of simulation. To illustrate the non negligible influence of the forces that are replaced by jumps, this figure also contains a radial distribution where long-range direct electrostatic interactions between $5$ and $9$ Angström are removed from the potential.
Table~\ref{table:1} shows a comparison in hydration free energy of a water molecule, of a sodium ion Na+ and of a potassium ion K+ in a system of $216$ water molecules simulated for $2ns$, using the Bennet Acceptance Ratio~\cite{BAR} (BAR) method. Note that the JUMP integrators are directly compatible with more advanced free energy methods~\cite{lagardere2024lambda}. The canonical measure is therefore similarly sampled by the JUMP integrators and the classical ones. Figure~\ref{fig:2} shows that the auto-diffusion constants (computed using the Einstein formula) are well preserved, thanks to the proximity of the process with the Langevin dynamics, as explained in Section~\ref{sec:proc-continu}. Again, the tests are done on a system of $500$ water molecules, simulated for $2ns$.

\paragraph{Resonance effects.} One of the well known issues of multi-timestep methods are resonances effects between internal periodicities of the molecular system and the non-physical periodicities created by the various time steps, that can induce numerical instabilities (such as abnormal fluctuations, or even explosion of the kinetic energy) and limit the largest time step that can be taken in the simulation. The reciprocal Coulomb part of the electrostatic interactions is partly responsible for those resonances (recall that this force is not purely long-range, since the erf function is not equal to zero around zero, and that the Ewald sum indeed implies all the atoms of the system). For this reason, increasing the real-space cutoff and replacing a part of the reciprocal electrostatic interactions by jumps, create much less oscillations in the system, thanks to the intrinsically random nature of velocity jumps, which allows to take a larger external time step in JUMP-RESPA1 than in BAOAB-RESPA1 and accelerate further the simulation.

\begin{figure}[htbp]
\begin{center}
    \subfloat[Various time steps for BAOAB-RESPA1]{\includegraphics[width=0.5\textwidth]{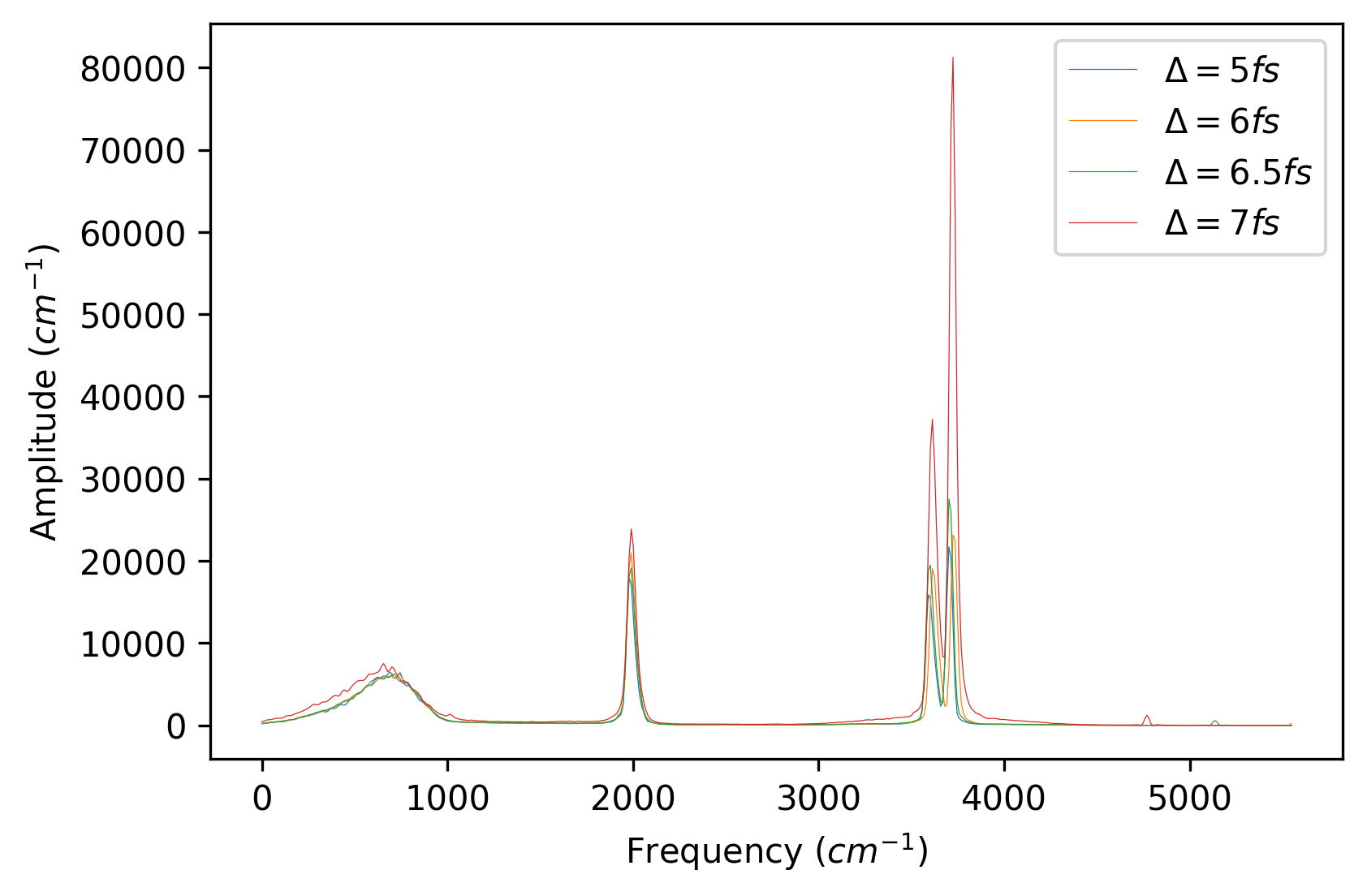}}
    \subfloat[Zoom on the non-physical artifacts]{\includegraphics[width=0.5\textwidth]{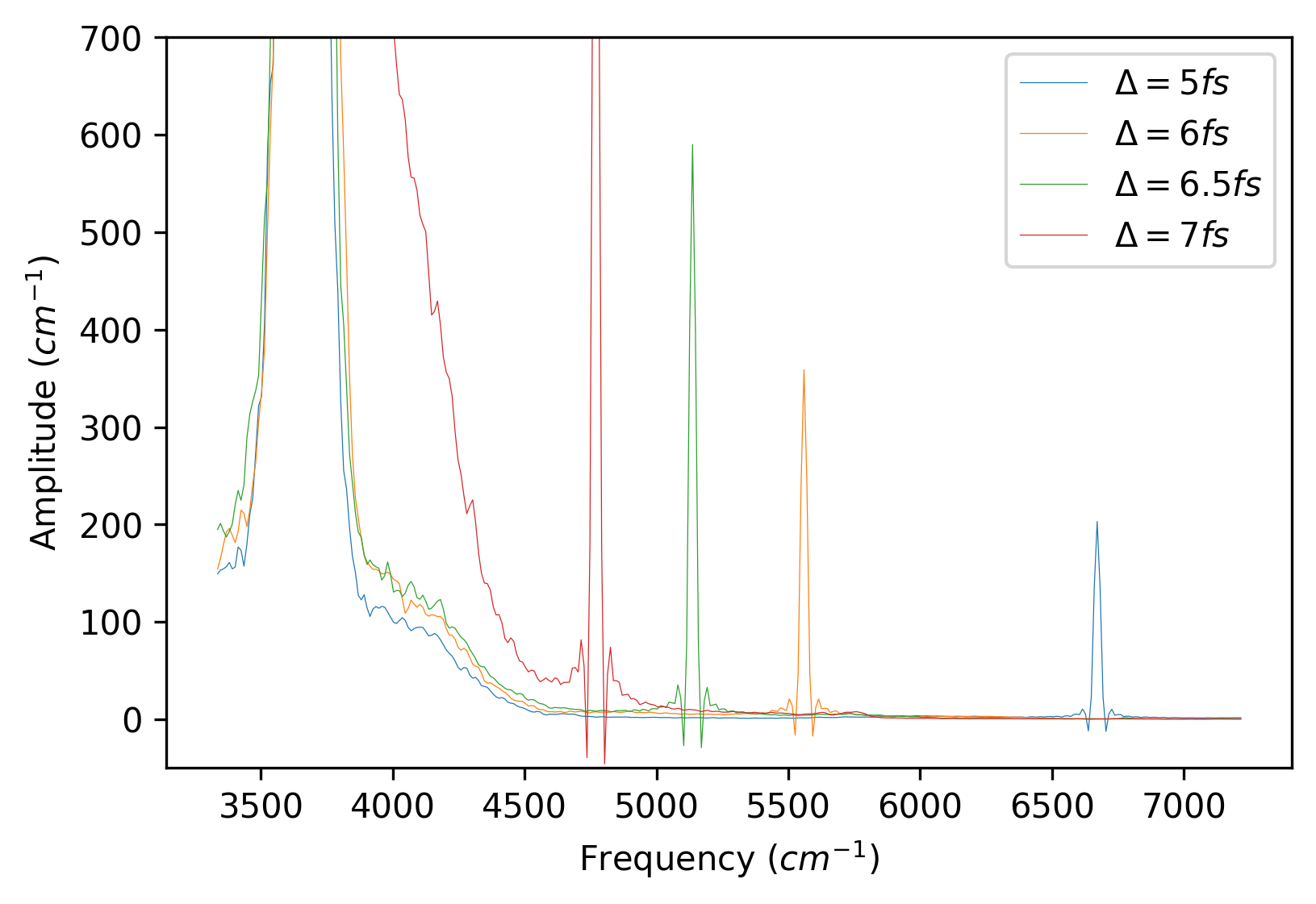}} \\
\end{center}
\caption{Velocity autocorrelation spectra of BAOAB-RESPA1 with various external time-steps, for a system of 500 water molecules. From left to right, the first bump corresponds to some many-body frequencies of the system, the first peak corresponds to intra-molecular angle oscillations, the second and third peaks correspond to bond oscillations. All the peaks at the right of those are non-physical artifacts due to the multi time-step. On the zoomed graph, the main non-physical peak moves to the left as the external time step increases, coupling with the bond-oscillation peak at the stability limit.}\label{fig:3}
\end{figure}

\begin{figure}[htbp]
\begin{center}
    \subfloat[$\Delta = 6fs$]{\includegraphics[width=0.5\textwidth]{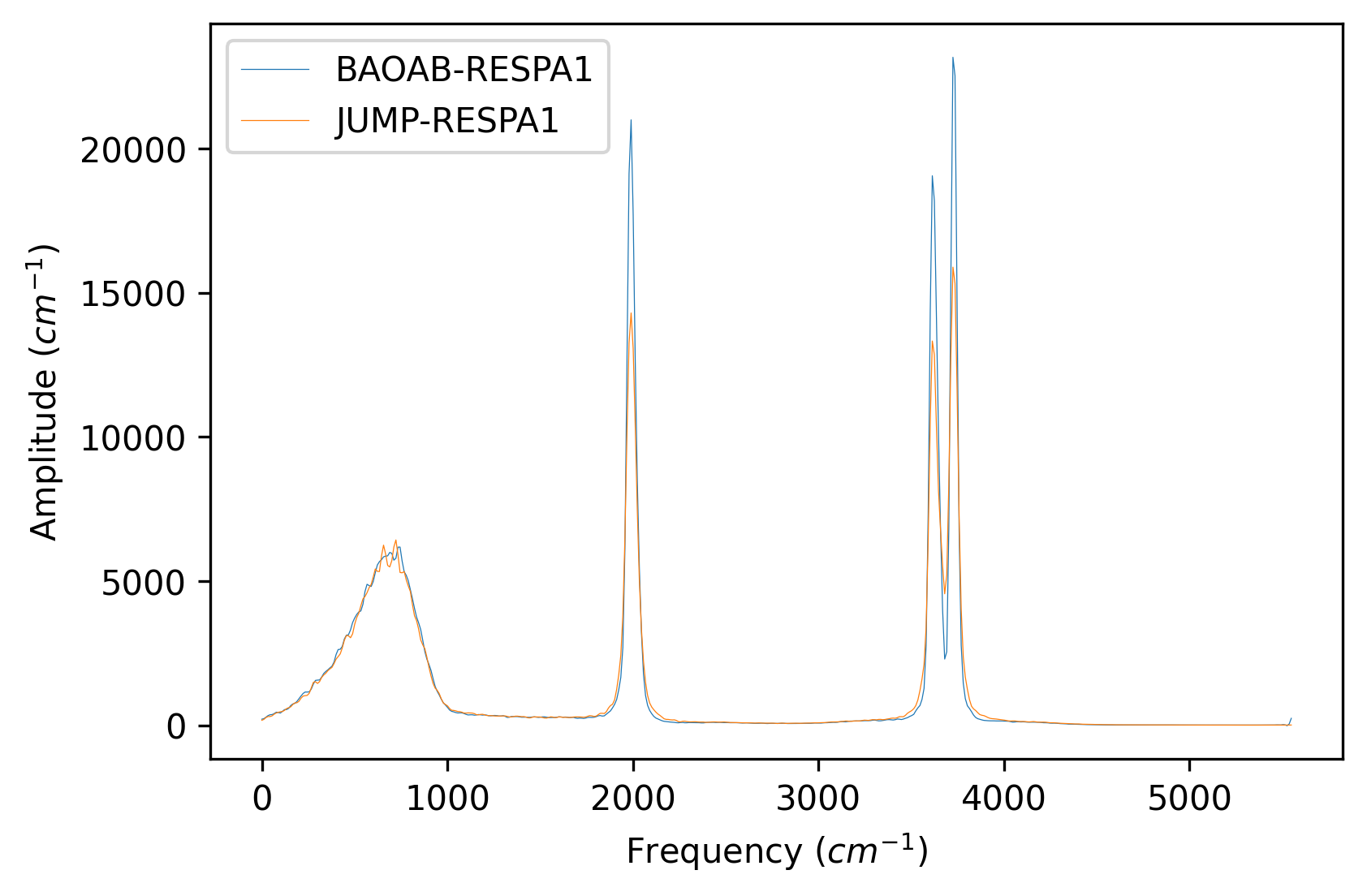}}
    \subfloat[Zoom on the non-physical artifacts, $\Delta = 6fs$ case]{\includegraphics[width=0.5\textwidth]{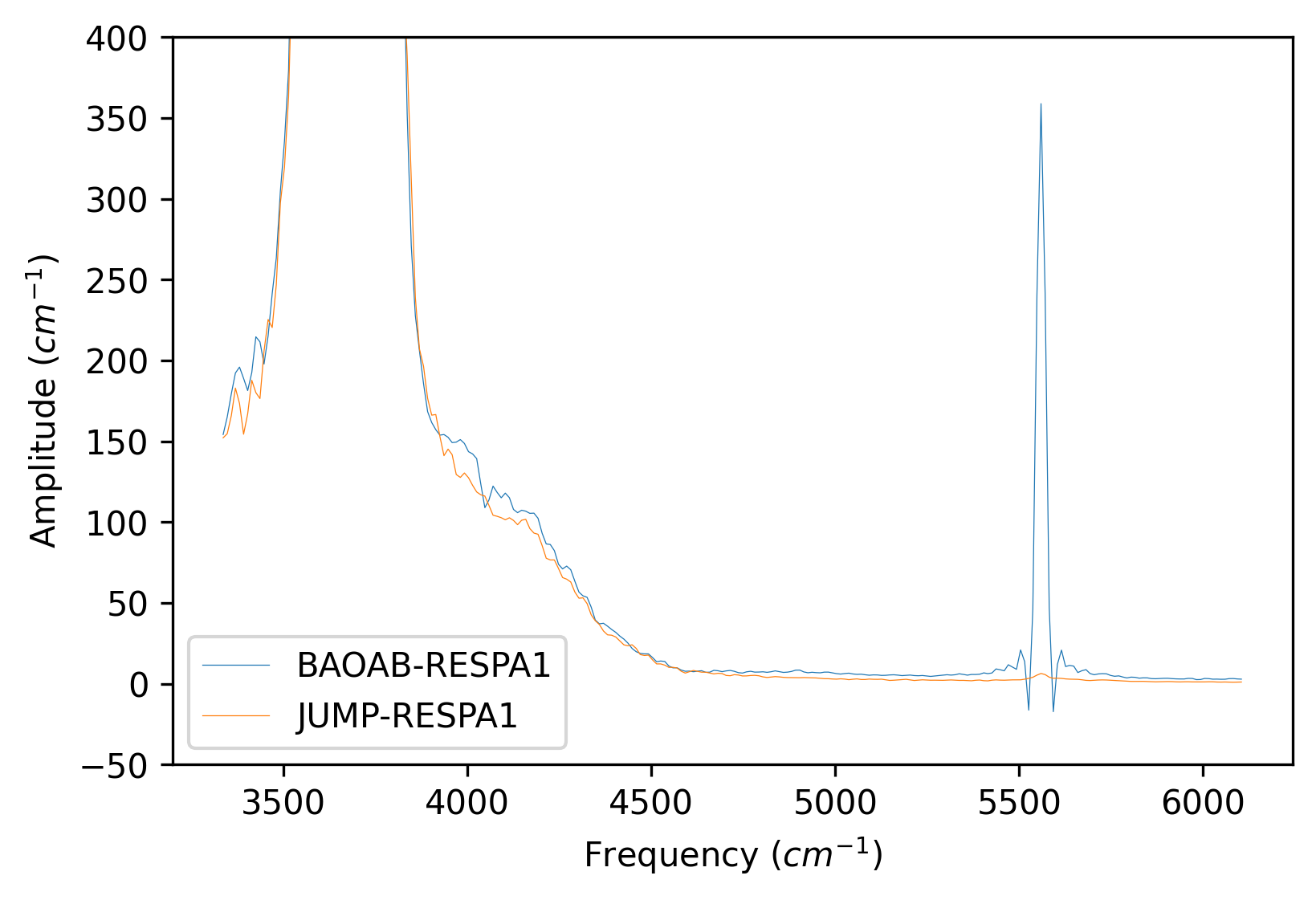}} \\
    \subfloat[$\Delta = 6.5fs$]{\includegraphics[width=0.5\textwidth]{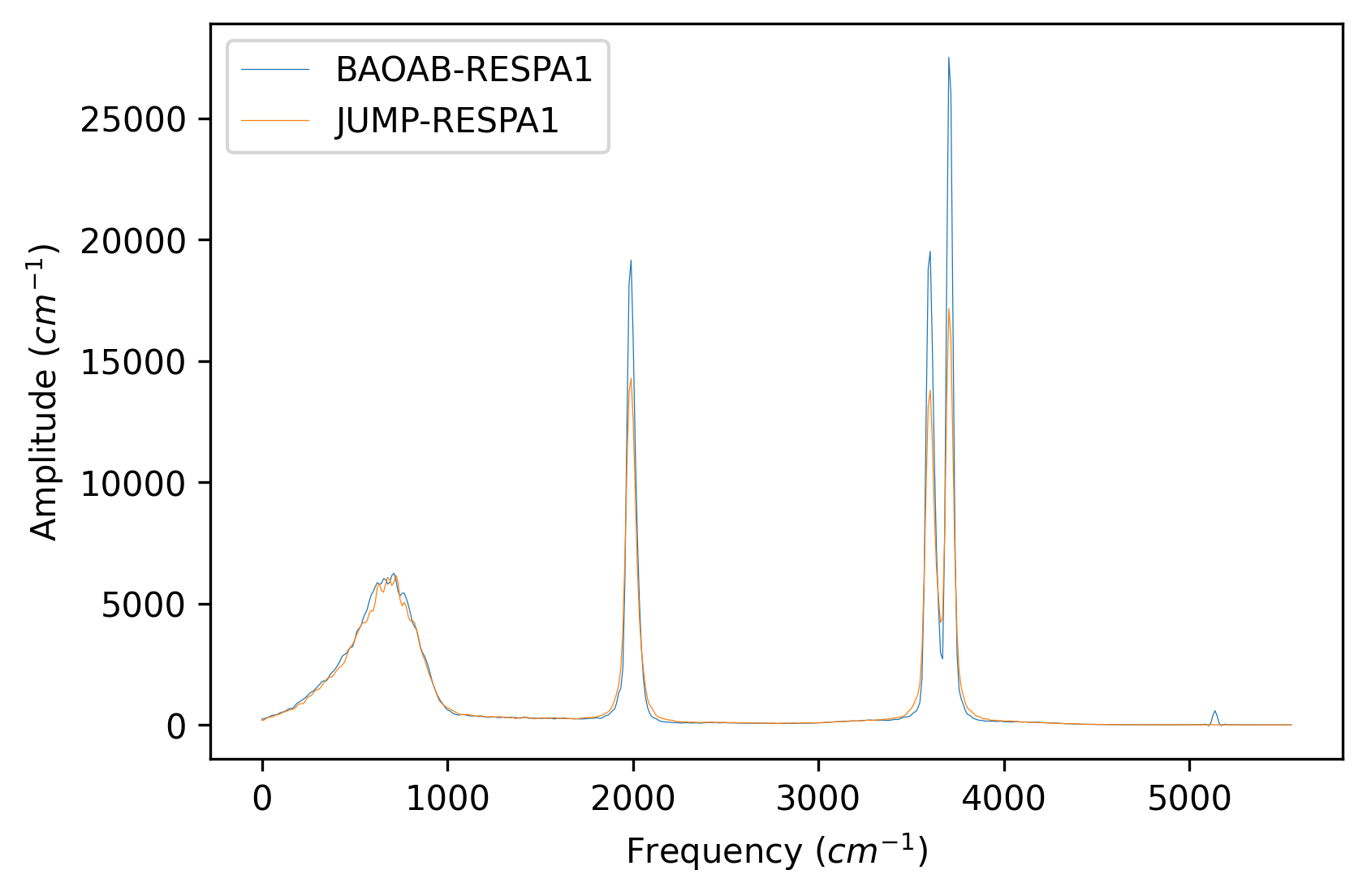}}
    \subfloat[Zoom on the non-physical artifacts, $\Delta = 6.5fs$ case]{\includegraphics[width=0.5\textwidth]{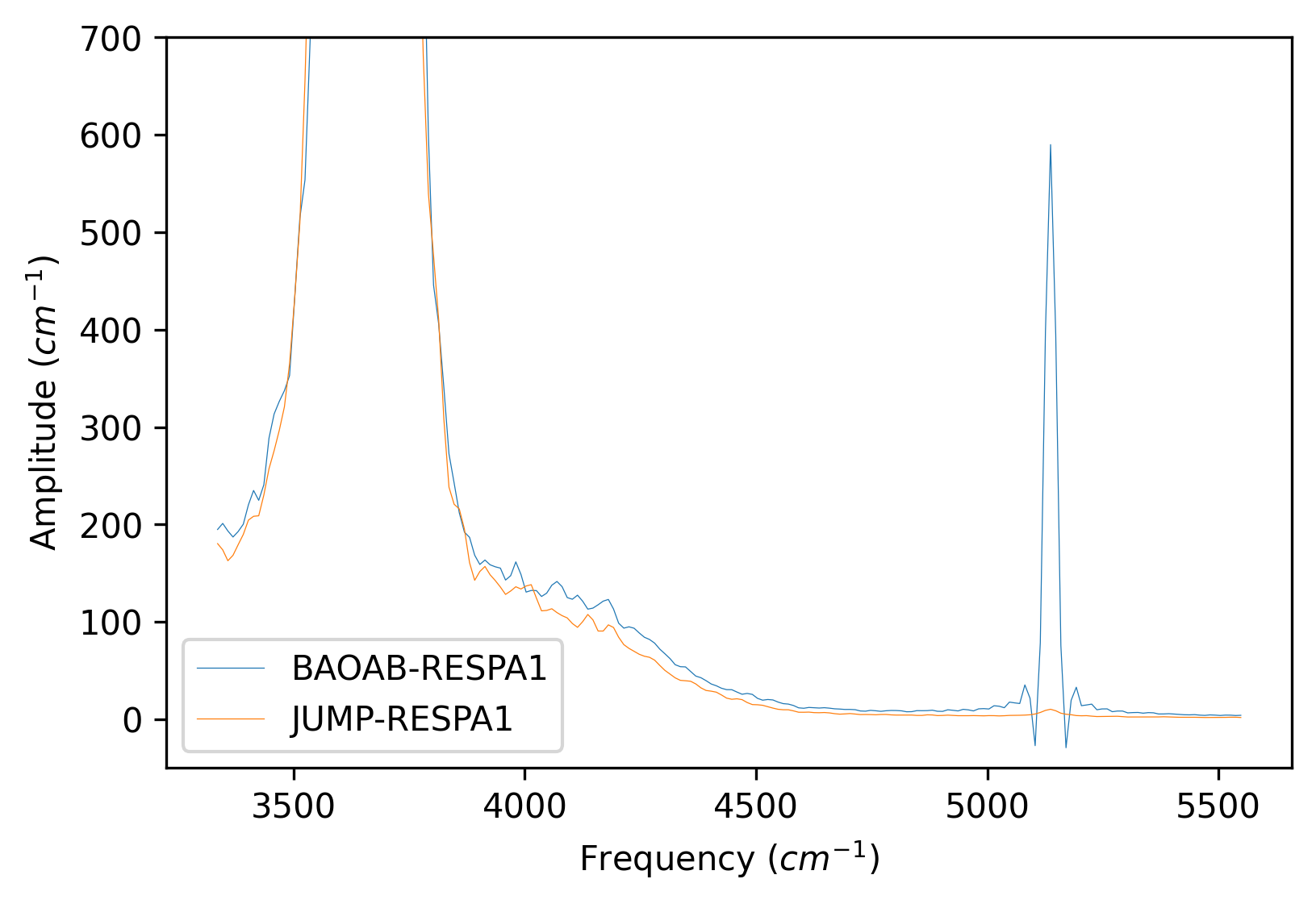}} \\
    \subfloat[$\Delta = 7fs$]{\includegraphics[width=0.5\textwidth]{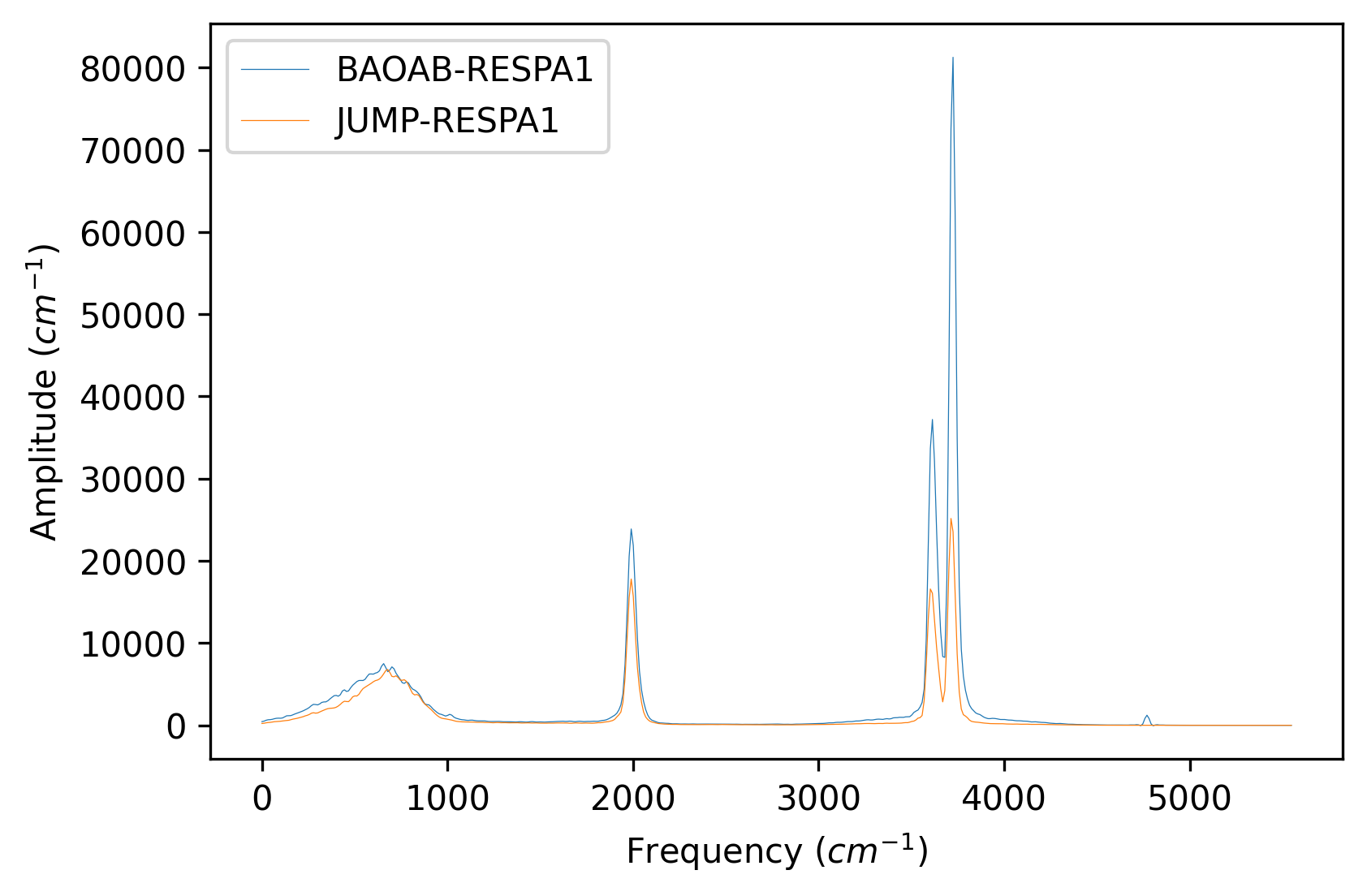}}
    \subfloat[Zoom on the non-physical artifacts, $\Delta = 7fs$ case]{\includegraphics[width=0.5\textwidth]{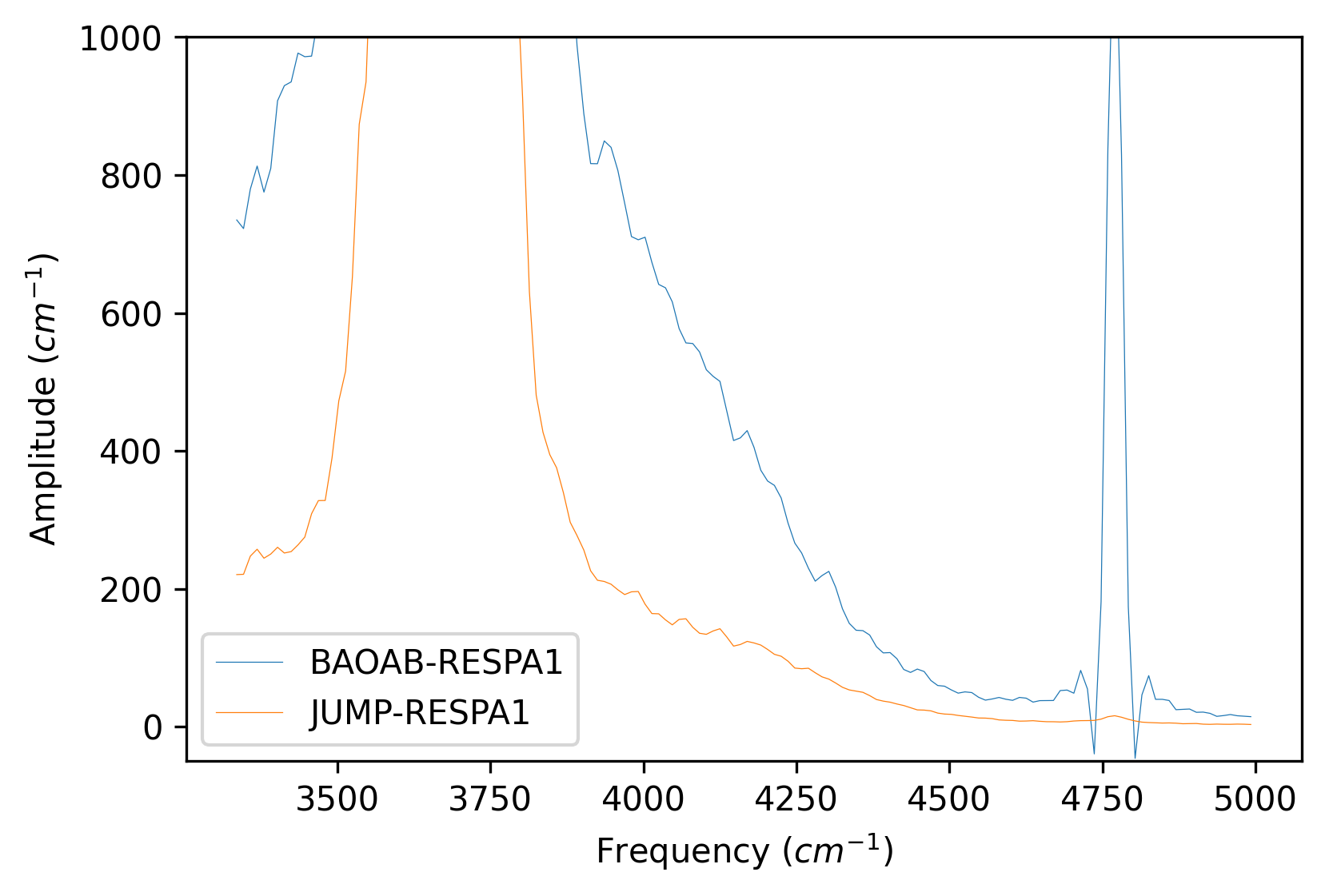}} \\
\end{center}
\caption{Velocity autocorrelation spectra of BAOAB-RESPA1 and JUMP-RESPA1 with various external time-steps. The non-physical artifacts are largely suppressed in JUMP-RESPA1.}\label{fig:4}
\end{figure}

These phenomena can be well observed on velocity autocorrelation spectra (computed from the MD trajectory at the level of the smallest time step using the implementation from ref.~\cite{ple2023routine}), illustrated in Figures~\ref{fig:3} and~\ref{fig:4}. On each graph, from left to right, the first band (below 1000 cm$^{-1}$) corresponds to molecule libration and slow fluctuations in the hydrogen bond network, the first peak (around 2000 cm$^{-1}$) corresponds to intra-molecular bending oscillations, the second and third peaks (around 3500 cm$^{-1}$) correspond to bond stretching motions. All the peaks at the right of those are non-physical artifacts due to the multi time-step. On the zoomed graphs, in BAOAB-RESPA1, the main non-physical peak moves to the left when the external time step increases, coming closer to the physical stretching peak, and even couples to it when the time step is too large. In JUMP-RESPA1, this phenomenon is largely suppressed, resulting in better numerical stability.

\paragraph{Performances of the algorithm.} In terms of simulation speed, the CPU version gives its best results on small systems ($35.5\%$ acceleration of JUMP with respect to BAOAB, and $21.6\%$ of acceleration of JUMP-RESPA1 with respect to BAOAB-RESPA1 on the system of $1500$ atoms), but this advantage disappears when looking at larger systems (with almost no difference between the algorithms when simulating $288000$ atoms). On the highly parallel GPU implementation however, it is the opposite: while there are almost no differences between JUMP and BAOAB on small systems (mostly because the GPU is not fully exploited on those, since the computational resources are not saturated, so tests on large molecular systems are more appropriate to this setting), the advantage appears on larger systems, and even is the best on the largest tested system ($21.3\%$ acceleration of JUMP with respect to BAOAB and $19.3\%$ acceleration of JUMP-RESPA1 with respect to BAOAB-RESPA1 on the system of $288000$ atoms). Table~\ref{table:2} shows the speed of the algorithms (in nanoseconds of simulation per day) on CPU and GPU implementations.

Those differences between GPU and CPU have different reasons. First, the neighbor lists are treated differently, see the end of this section for more details. Then, the proportion of time spent in the different routines of the integrator is different on CPU and GPU.
For instance, the parallel architecture of GPUs allows for a more efficient treatment of the van der Waals and direct electrostatic interactions, while the many-body Ewald sum takes almost $50\%$ of the total computation time, especially on larger systems. As a consequence, on a large system, reducing the reciprocal part of electrostatics and treating a larger portion of the potential with jumps yields a significant computational advantage. For reference, Table~\ref{table:4} gives the proportion of time spent in each of the main routines of the CPU and GPU implementations, comparing two system sizes.

As explained in~\cite{GROMACS}, when increasing the real-space cutoff $r_c$, the Fourier space spacing can be multiplied by the same factor, which reduces the cost of its computation. It is particularly advantageous to do that on the large systems, where the interpolation of the structure factors of the Ewald sum is done on large grids.

\begin{table}[htbp]
\begin{center}
\begin{tabular}{|l|l|l|}
\hline
Integrator                     & BAOAB & JUMPS         \\ \hline
$\Delta G_{\text{hydrat}}$ $\text{H}_2\text{O}$ ($Kcal/Mol$) & $6.73 \pm 0.09$ & $6.77 \pm 0.08$ \\ \hline
$\Delta G_{\text{hydrat}}$ Na+ ($Kcal/Mol$) & $75.48 \pm 0.09$ & $75.61 \pm 0.03$ \\ \hline
$\Delta G_{\text{hydrat}}$ K+ ($Kcal/Mol$) & $58.81 \pm 0.04$ & $58.82 \pm 0.02$ \\ \hline
\end{tabular}
\end{center}
\caption{Hydration free energies of water, sodium and potassium in $216$ water molecules.}
\label{table:1}
\end{table}

\begin{table}[htbp]
\begin{center}
\begin{tabular}{ll|llll|llll|}
\cline{3-10}
\multicolumn{2}{l|}{}                                                             & \multicolumn{4}{l|}{CPU}                                                                                                                                                                                                                                             & \multicolumn{4}{l|}{GPU}                                                                                                                                                                                                                                              \\ \hline
\multicolumn{2}{|l|}{Number of atoms}                                             & \multicolumn{1}{l|}{1500}                                                  & \multicolumn{1}{l|}{12000}                                                 & \multicolumn{1}{l|}{96000}                                                 & 288000                        & \multicolumn{1}{l|}{1500}                                                   & \multicolumn{1}{l|}{12000}                                                 & \multicolumn{1}{l|}{96000}                                                 & 288000                        \\ \hline
\multicolumn{2}{|l|}{BAOAB (1 fs)}                                                & \multicolumn{1}{l|}{3.50}                                                  & \multicolumn{1}{l|}{3.95e-1}                                               & \multicolumn{1}{l|}{3.89e-2}                                               & 8.15e-3                       & \multicolumn{1}{l|}{178.0}                                                  & \multicolumn{1}{l|}{98.19}                                                 & \multicolumn{1}{l|}{16.91}                                                 & 5.13                          \\ \hline
\multicolumn{2}{|l|}{JUMP (1 fs)}                                                 & \multicolumn{1}{l|}{4.74}                                                  & \multicolumn{1}{l|}{4.94e-1}                                               & \multicolumn{1}{l|}{4.52e-2}                                               & 8.74e-3                       & \multicolumn{1}{l|}{175.3}                                                  & \multicolumn{1}{l|}{1.7.69}                                                & \multicolumn{1}{l|}{20.14}                                                 & 6.22                          \\ \hline
\rowcolor[HTML]{C0C0C0} 
\multicolumn{2}{|l|}{\cellcolor[HTML]{C0C0C0}{\color[HTML]{000000} Acceleration}} & \multicolumn{1}{l|}{\cellcolor[HTML]{C0C0C0}{\color[HTML]{000000} 35.5\%}} & \multicolumn{1}{l|}{\cellcolor[HTML]{C0C0C0}{\color[HTML]{000000} 25.1\%}} & \multicolumn{1}{l|}{\cellcolor[HTML]{C0C0C0}{\color[HTML]{000000} 16.2\%}} & {\color[HTML]{000000} 7.24\%} & \multicolumn{1}{l|}{\cellcolor[HTML]{C0C0C0}{\color[HTML]{000000} -1.52\%}} & \multicolumn{1}{l|}{\cellcolor[HTML]{C0C0C0}{\color[HTML]{000000} 9.68\%}} & \multicolumn{1}{l|}{\cellcolor[HTML]{C0C0C0}{\color[HTML]{000000} 19.1\%}} & {\color[HTML]{000000} 21.3\%} \\ \hline
\multicolumn{2}{|l|}{BAOAB-RESPA (1:2 fs)}                                        & \multicolumn{1}{l|}{5.68}                                                  & \multicolumn{1}{l|}{6.29e-1}                                               & \multicolumn{1}{l|}{6.25e-2}                                               & 1.46e-2                       & \multicolumn{1}{l|}{158.4}                                                  & \multicolumn{1}{l|}{110.78}                                                & \multicolumn{1}{l|}{27.61}                                                 & 8.59                          \\ \hline
\multicolumn{2}{|l|}{JUMP-RESPA (1:2 fs)}                                         & \multicolumn{1}{l|}{6.62}                                                  & \multicolumn{1}{l|}{6.98e-1}                                               & \multicolumn{1}{l|}{6.24e-2}                                               & 1.38e-2                       & \multicolumn{1}{l|}{154.7}                                                  & \multicolumn{1}{l|}{115.93}                                                & \multicolumn{1}{l|}{30.58}                                                 & 9.82                          \\ \hline
\rowcolor[HTML]{C0C0C0} 
\multicolumn{2}{|l|}{\cellcolor[HTML]{C0C0C0}Acceleration}                        & \multicolumn{1}{l|}{\cellcolor[HTML]{C0C0C0}16.5\%}                        & \multicolumn{1}{l|}{\cellcolor[HTML]{C0C0C0}11.0\%}                        & \multicolumn{1}{l|}{\cellcolor[HTML]{C0C0C0}-0.16\%}                       & -5.48\%                       & \multicolumn{1}{l|}{\cellcolor[HTML]{C0C0C0}-2.34\%}                        & \multicolumn{1}{l|}{\cellcolor[HTML]{C0C0C0}4.65\%}                        & \multicolumn{1}{l|}{\cellcolor[HTML]{C0C0C0}10.8\%}                        & 14.3\%                        \\ \hline
\multicolumn{2}{|l|}{BAOAB-RESPA1 (1:2:5 fs)}                                     & \multicolumn{1}{l|}{6.75}                                                  & \multicolumn{1}{l|}{7.13e-1}                                               & \multicolumn{1}{l|}{6.73e-2}                                               & 1.82e-2                       & \multicolumn{1}{l|}{276.6}                                                  & \multicolumn{1}{l|}{141.3}                                                 & \multicolumn{1}{l|}{34.69}                                                 & 11.93                         \\ \hline
\multicolumn{2}{|l|}{JUMP-RESPA1 (1:2:6 fs)}                                      & \multicolumn{1}{l|}{8.21}                                                  & \multicolumn{1}{l|}{8.30e-1}                                               & \multicolumn{1}{l|}{6.58e-2}                                               & 1.81e-2                       & \multicolumn{1}{l|}{308.9}                                                  & \multicolumn{1}{l|}{158.6}                                                 & \multicolumn{1}{l|}{41.48}                                                 & 14.23                         \\ \hline
\rowcolor[HTML]{C0C0C0} 
\multicolumn{2}{|l|}{\cellcolor[HTML]{C0C0C0}Acceleration}                        & \multicolumn{1}{l|}{\cellcolor[HTML]{C0C0C0}21.6\%}                        & \multicolumn{1}{l|}{\cellcolor[HTML]{C0C0C0}16.4\%}                        & \multicolumn{1}{l|}{\cellcolor[HTML]{C0C0C0}-2.22\%}                       & -0.55\%                       & \multicolumn{1}{l|}{\cellcolor[HTML]{C0C0C0}11.7\%}                         & \multicolumn{1}{l|}{\cellcolor[HTML]{C0C0C0}12.2\%}                        & \multicolumn{1}{l|}{\cellcolor[HTML]{C0C0C0}19.6\%}                        & 19.3\%                        \\ \hline
\end{tabular}
\end{center}
\caption{Speed performances of the algorithms, expressed in nanoseconds of simulation per day}
\label{table:2}
\end{table}

\begin{table}[htbp]
\begin{center}
\begin{tabular}{l|lll|llll|}
\cline{2-8}
                                  & \multicolumn{3}{l|}{BAOAB on CPU}                                            & \multicolumn{4}{l|}{JUMP on CPU}                                                                          \\ \hline
\multicolumn{1}{|l|}{Nb of atoms} & \multicolumn{1}{l|}{vdW}     & \multicolumn{1}{l|}{Direct Coulomb} & SPME    & \multicolumn{1}{l|}{vdW}    & \multicolumn{1}{l|}{Direct Coulomb} & \multicolumn{1}{l|}{SPME}    & Jumps  \\ \hline
\multicolumn{1}{|l|}{12000}       & \multicolumn{1}{l|}{8.34\%}  & \multicolumn{1}{l|}{41.23\%}        & 37.44\% & \multicolumn{1}{l|}{5.24\%} & \multicolumn{1}{l|}{21.79\%}        & \multicolumn{1}{l|}{43.63\%} & 2.41\% \\ \hline
\multicolumn{1}{|l|}{288000}      & \multicolumn{1}{l|}{13.64\%} & \multicolumn{1}{l|}{33.25\%}        & 40.45\% & \multicolumn{1}{l|}{7.71\%} & \multicolumn{1}{l|}{38.48\%}        & \multicolumn{1}{l|}{23.20\%} & 3.37\% \\ \hline
\multicolumn{1}{|l|}{}            & \multicolumn{3}{l|}{BAOAB on GPU}                                            & \multicolumn{4}{l|}{JUMP on GPU}                                                                          \\ \hline
\multicolumn{1}{|l|}{12000}       & \multicolumn{1}{l|}{25.5\%}  & \multicolumn{1}{l|}{34.0\%}         & 17.7\%  & \multicolumn{1}{l|}{26.2\%} & \multicolumn{1}{l|}{21.9\%}         & \multicolumn{1}{l|}{18.6\%}  & 8.0\%  \\ \hline
\multicolumn{1}{|l|}{280000}      & \multicolumn{1}{l|}{8.5\%}   & \multicolumn{1}{l|}{32.3\%}         & 52.0\%  & \multicolumn{1}{l|}{6.6\%}  & \multicolumn{1}{l|}{46.7\%}         & \multicolumn{1}{l|}{28.0\%}  & 7.4\%  \\ \hline
\end{tabular}
\end{center}
\caption{Time spent in each routine}
\label{table:4}
\end{table}

\paragraph{Simulation details}

\begin{itemize}
\item Tinker-HP version: 1.2
\item Force-field: OPLS-AA
\item Water model: SPC/Fw
\item switching parameter: $0.5$Å
\item Friction: $1.0ps^{-1}$
\item Temperature: $300 K$.
\end{itemize}

Parameters for BAOAB, BAOAB-RESPA and BAOAB-RESPA1 simulations:

\begin{itemize}
    \item Time step for BAOAB: $\delta = 1fs$.
    \item Time steps for BAOAB-RESPA: $\delta = 1fs$, $\Delta = 2fs$.
    \item Time steps for BAOAB-RESPA1: $\delta = 1fs$, $\kappa = 2fs$, $\Delta = 5fs$.
    \item Charge real-space cutoff: $7$Å.
    \item Ewald parameter: $\alpha = 0.5446$.
    \item Van de Waals long range cutoff: $10$Å.
\end{itemize}

Parameters for JUMP, JUMP-RESPA and JUMP-RESPA1 simulations:

\begin{itemize}
    \item Time step for JUMP: $\delta = 1fs$.
    \item Time steps for JUMP-RESPA: $\delta = 1fs$, $\Delta = 2fs$.
    \item Time steps for JUMP-RESPA1: $\delta = 1fs$, $\kappa = 2fs$, $\Delta = 6.5fs$.
    \item Van der Waals jumps only on CPU.
    \item Van de Waals long range cutoff: $10$Å.
    \item Van de Waals short range cutoff (if activated): $7$Å.
    \item Type of jump parameter: adaptive $\veps$.
\end{itemize}

\begin{table}[htbp]
\begin{center}
\begin{tabular}{|l|l|l|l|l|l|}
\hline
Number of atoms                  & 648      & 1500     & 12000    & 96000       & 288000      \\ \hline
Box size (in Å)                  & 18.643   & 24.662   & 49.232   & 98.5        & 142.27      \\ \hline
default PME-grid                 & 24x24x24 & 30x30x30 & 60x60x60 & 120x120x120 & 180x180x180 \\ \hline
PME-grid in JUMP                 & 24x24x24 & 30x30x30 & 60x60x60 & 96x96x96    & 128x128x128 \\ \hline
$\alpha$ parameter in JUMP       & 0.554    & 0.554    & 0.554    & 0.42        & 0.3767      \\ \hline
Real-space cutoff JUMP (in Å)    & 7        & 7        & 7        & 9           & 10          \\ \hline
Short-range charge cutoff (in Å) & 5        & 5        & 5        & 6           & 7           \\ \hline
Jump parameter $\veps_0$                        & 0.1      & 0.1      & 0.1      & 0.01        & 0.01        \\ \hline
\end{tabular}
\end{center}
\caption{JUMP and PME-grid parameters.}
\end{table}

\section{Conclusion and perspectives}

To conclude, our work presents a new framework to construct velocity jumps (or JUMP) integrators for molecular dynamics simulations that can be parallelized on GPU and integrated into multi-timestep methods. It allows to accelerate simulations while preserving sampling precision and dynamical properties, and avoiding certain classical resonance issues. The JUMP approaches provide significant acceleration over their BAOAB, BAOAB-RESPA and BAOAB-RESPA1 counterparts. If the present implementation has been performed in the Tinker-HP package, the approach could be easily ported to other molecular dynamics software. Future work will deal with the extension of the approach to polarizable force fields \cite{article69,article70,article71,article72} and with its combination with advanced periodic boundary conditions treatment of electrostatics such as ANKH \cite{ANKH}.

\paragraph{Supporting information} The supporting information contains rigorous definitions on the continuous-time process and splitting schemes, proofs of the mathematical statements as well as technical details on the TinkerHP implementation of the algorithms.

\paragraph{Conflicts of interest}

Louis Lagardère and Jean-Philip Piquemal are co-founders and shareholders of Qubit Pharmaceuticals. The remaining authors declare no competing interests.

\paragraph{Acknowledgments} 
This work was made possible thanks to funding from the European Research Council (ERC) under the European Union's Horizon 2020 research and innovation program (grant agreement No 810367), project EMC2 (JPP). Computations have been performed at IDRIS, GENCI (Jean Zay machine, France) on grant no A0130712052 (JPP).
\pagebreak
\bibliographystyle{unsrt}
\bibliography{biblio}

\end{document}